\documentclass[11pt,a4paper]{article}
\usepackage[top=1in, bottom=1in, left=1in, right=1in]{geometry}
\usepackage{bm}
\usepackage{natbib}
\usepackage{tabularx}
\usepackage{graphicx,lscape}
\usepackage{caption}
\usepackage{subcaption}
\usepackage{booktabs}
\usepackage{threeparttable}
\usepackage{enumitem}
\usepackage{amsmath,amsthm,amssymb}
\usepackage{xcolor}
\usepackage{makecell} 
\usepackage{algorithm}
\usepackage{algorithmic}
\usepackage{comment}

\usepackage{longtable}
\usepackage[colorlinks=true,citecolor=blue]{hyperref}
\usepackage{cleveref}

\newcommand{\MBA}{\mathbf{A}}

\newcommand{\MBF}{\mathbf{F}}

\newcommand{\MBI}{\mathbf{I}}

\newcommand{\MBy}{\mathbf{y}}

\newcommand{\bed}{\begin{displaymath}}
\newcommand{\eed}{\end{displaymath}}
\newcommand{\bel}{\begin{eqalign}}
\newcommand{\eel}{\end{eqalign}}
\newcommand{\bee}{\begin{equation}}
\newcommand{\eee}{\end{equation}}

\newcommand*\oline[1]{%
  \vbox{%
    \hrule height 0.1pt
    \kern0.25ex
    \hbox{%
      \kern-0.1em
      \ifmmode#1\else\ensuremath{#1}\fi
      \kern-0.1em
    }
  }
}
\theoremstyle{definition}

\newtheorem{assump}{Assumption}

\def\blue#1{{\color{blue}#1}}

\title{\bfseries Spending Behavior and Economic Impacts of Urban Digital Consumption Vouchers}
\author{
  Ming-Huan Liou\thanks{\blue{The authors gratefully acknowledge the Taipei City Office of Commerce, Taipei City Government, for providing access to the user survey data collected through the commissioned project ``Usage Survey and Effectiveness Evaluation of Taipei Voucher 2.0'' (Project No. 11202), which served as the foundation for the analysis conducted in this study.}} \\
  \normalsize Taiwan Institute of Economic Research
  \and
  Shou-Yung Yin \\
  \normalsize National Taipei University
  \and
  Hsiang-Wen Mao \\
  \normalsize National Taiwan University
}
\date{This Version: \today}

\begin{document}
\thispagestyle{empty}

\maketitle

\vspace{0.5cm}
\begin{center}
  {\large\bf Abstract}
\end{center}
\vspace{-0.2cm}

\noindent
This paper evaluates the Taipei Bear Vouchers 2.0 program using verified user-level survey data and a regional input–output model to assess the effectiveness of consumption vouchers as a fiscal stimulus tool. We focus on three key behavioral mechanisms: expenditure substitution, induced consumption, and the intensity of treatment through varying voucher face values. Our findings show that voucher effectiveness differs by type. Accommodation vouchers stimulate the most additional spending due to low expenditure substitution and high induced consumption effects, while sports vouchers often replace existing consumption. Increases in voucher value further enhance marginal consumption, especially when this change is a part of unexpected policy. Taking these behavioral responses into account, we find that the output multiplier of the program rises significantly, and indirect benefits extend to untargeted sectors through inter-industry linkages. These results highlight the critical role of consumer behavior in shaping policy outcomes and offer practical guidance for designing more effective and targeted consumption voucher programs.

\vspace{1cm}


\vfill

\setcounter{page}{1}
\section{Introduction}
The COVID-19 pandemic triggered one of the most severe global economic shocks in recent history, causing widespread disruptions to labor markets, consumer spending, and industrial production. In response, governments around the world adopted a range of expansionary fiscal measures to stimulate economic recovery \citep{IMF2022}. Among these policies, two broad approaches emerged: emergency relief, which provided financial assistance to businesses and households experiencing operational difficulties, and economic stimulus, which distributed cash payments or consumption vouchers to households. While the former focused on maintaining economic stability, the latter aimed to actively boost consumption and revive demand. Accordingly, the specific mix and design of these interventions varied considerably across countries.

Several countries implemented large-scale fiscal interventions in response to the COVID-19 pandemic. In the United States, the Coronavirus Aid, Relief, and Economic Security (CARES) Act, enacted in 2020, amounted to a USD 2.2 trillion stimulus package, which was the largest in U.S. history. It included direct cash payments to individuals, expanded unemployment insurance, and sector-specific financial support. Japan introduced three rounds of stimulus packages totaling more than 300 trillion yen, equivalent to approximately 60 percent of its GDP. These measures featured universal cash payments along with subsidies for fuel and utility expenses. South Korea also provided direct cash payments to all households, and Singapore launched the Care and Support Package to deliver one-time cash transfers. In contrast, Taiwan not only launched general-purpose consumption vouchers, as in other countries, but also adopted a more targeted strategy by issuing vouchers with restricted usage conditions, aiming to encourage spending in specific sectors that were most severely affected by the pandemic.\footnote{In Taiwan, the government also introduced a relief and revitalization plan for small and medium-sized enterprises, the effects of which have been analyzed in \citet{chen2025pandemic}.}

These policy approaches raise an important question: when the objective is to stimulate private consumption, should governments prioritize direct cash transfers or the issuance of consumption vouchers? The economic impacts and trade-offs associated with these two instruments have been widely discussed in the literature. Cash transfers offer two primary advantages. First, they provide recipients with maximum flexibility, allowing individuals to decide whether to spend, save, or repay debt. Second, they generally entail lower administrative costs than the printing and distribution of paper-based vouchers. However, this flexibility may reduce their stimulative effect, as many households choose to save the funds or pay down debt rather than increase consumption \citep{coibion2020did}. Additionally, cash transfers cannot be easily directed toward specific sectors that are most severely affected during downturns \citep{kim2024can}.

In contrast, consumption vouchers may help preserve the intended stimulative effect by limiting flexibility and channeling spending toward targeted sectors. While \citet{kan2017understanding} reported limited effects from Taiwan's 2008 paper-based voucher program, \citet{xing2023quick} and \citet{chen2025revenue} documented strong stimulus outcomes associated with digital consumption vouchers in China. Moreover, the latter two studies highlight several advantages of digital voucher schemes, including lower administrative costs, faster distribution, real-time tracking of transactions, and the ability to support specific industries such as retail and food services.

Nevertheless, the existing literature on voucher usage and effectiveness faces two main limitations. First, most studies rely on telephone surveys, which make it difficult to verify the authenticity of responses and whether the vouchers were actually used (e.g., \citet{kan2017understanding}). Second, much of the literature focuses on the effects of a single voucher type, without investigating how different consumer groups respond to various types of vouchers (e.g., \citet{leone1996coupon,xing2023quick,chen2025revenue}).

To address these gaps, this study examines the Taipei Bear Vouchers 2.0 program, launched by the Taipei City Government in October 2022, as a case study. Using first-hand survey data collected through the TaipeiPASS platform, we investigate how different types of vouchers affect spending behavior and how consumers with different demographic characteristics respond to the program. The use of verified and platform-based data allows for greater confidence in the accuracy and authenticity of voucher usage.

More specifically, this study aims to identify three key policy-relevant effects: the expenditure substitution effect, the induced consumption effect, and the intensity of treatment associated with varying face values of vouchers. These effects are analyzed across different types of vouchers. The expenditure substitution effect arises when consumers use vouchers to purchase goods and services that they would have bought regardless, treating the vouchers as a substitute for cash. In contrast, the induced consumption effect refers to additional spending that exceeds the face value of the voucher when consumers redeem them. The intensity of treatment captures the marginal effect of increasing the face value of vouchers, helping to quantify how much additional spending is generated by higher-value instruments. Understanding these effects is of critical importance to policymakers. If the goal is to stimulate consumption and maximize the fiscal multiplier through voucher programs, priority should be given to designing and allocating vouchers that minimize substitution and enhance induced consumption. To further evaluate the broader economic impact of voucher-induced spending, we follow the methodological approaches proposed by \citet{hua2022evaluating} and \citet{chen2016modeling}, and apply a regional input–output model to estimate sector-specific output multipliers.

In addition to identifying these effects, we also address the issue of self-reporting bias, which is a common concern in survey-based studies \citep{geisen2012examining}. To account for this, we adopt a conservative approach by treating the maximum reported value within a finely stratified demographic subgroup as the upper bound of potential bias. We then construct two confidence intervals using a stratified bootstrap procedure: one based on estimates without bias correction, and another based on the most conservative scenario. This dual-interval approach yields a comprehensive confidence region that ensures the robustness of our estimates. These bias-correction considerations and the associated inference framework contribute a novel robustness feature to the existing literature on consumption vouchers.

This paper reports three main findings. First, expenditure substitution rates vary significantly across voucher types. Sports vouchers exhibit the highest substitution rates, suggesting that spending in this category often replaces pre-planned consumption, likely due to the habitual nature of sports-related purchases. In contrast, accommodation vouchers show the lowest substitution rates, indicating greater potential to generate new spending. Second, induced consumption effects are substantial. The accommodation voucher again stands out, with the highest induced consumption, reinforcing its effectiveness in promoting incremental and geographically distributed spending. Third, regional input–output analysis reveals that the baseline output multiplier of the Taipei Bear Vouchers 2.0 program is 0.97 when behavioral responses are not considered. Once substitution and induced consumption effects are incorporated, the multiplier rises to as high as 1.76, indicating that each NT\$1 in voucher spending can generate up to NT\$1.76 in regional economic output. These findings highlight the critical role of consumer behavior in shaping the effectiveness of voucher-based fiscal stimulus and underscore the importance of careful program design tailored to spending elasticity and sectoral targeting. Analysis of treatment intensity further shows that even modest increases in voucher face value can generate meaningful marginal spending effects, especially when such increases are unexpected or paired with vendor promotions.

The remainder of this paper is organized as follows: Section 2 analyzes
the related literature; Section 3 provides background information and
describes the data; Section 4 outlines the research method; Section 5
presents the empirical results; and Section 6 concludes our findings.

\section{Literature Review}
\subsection{Direct Cash Transfer}
During economic downturns, fiscal stimulus measures aimed at enhancing
consumer demand have consistently served as essential tools for
policymakers. Instruments such as direct cash transfers and tax cuts
have been extensively utilized to boost aggregate demand by increasing
household disposable income. However, empirical research has highlighted
the limitations of cash-based stimulus programs in generating strong
multiplier effects. Studies by \citet{shapiro2003consumer,shapiro2009did}, \citet{johnson2006household}, and \citet{parker2013consumer} indicate that the marginal
propensity to consume (MPC) from such transfers often falls between 0.2
and 0.9, implying that a significant portion of the funds does not
immediately translate into consumption. \citet{coibion2020did} found that
only about 40\% of households used the CARES Act cash transfers for
consumption, while approximately 60\% either saved the funds or used
them to reduce debt.

\subsection{Consumption Vouchers}

In addition to direct cash transfers or tax cuts, consumption vouchers
have also been widely used by governments as a fiscal tool to stimulate
demand during economic downturns. Early voucher programs were
distributed in paper form, which incurred relatively high administrative
costs. Furthermore, due to the lack of strict usage restrictions, these
vouchers were often used as a substitute for cash, thereby limiting
their effectiveness in generating additional consumption as intended. For example, \citet{kan2017understanding} found a marginal propensity to consume of 0.243 when analyzing Taiwan's 2008 paper-based voucher program, indicating limited economic stimulus, as most recipients used the vouchers for planned rather than additional spending.

To address the limited economic effectiveness of cash transfers and the
additional costs associated with paper-based consumption vouchers,
digital coupons have emerged as a novel and technologically enabled
fiscal tool. By setting minimum spending thresholds and expiration
dates, digital coupons are designed to encourage immediate consumption
and can be strategically directed at affected sectors such as food
services and retail \citep[e.g.,][]{leone1996coupon,delvecchio2006effect}.  Advances in mobile payment technology, especially QR code systems, have further enhanced the efficiency and scalability of coupon
distribution \citep{agarwal2020fintech}.

\citet{liu2021stimulating} examined a large-scale digital coupon program in
Hangzhou, China, and found that it successfully boosted immediate
consumer spending, suggesting broad applicability for local economic
recovery efforts. Similarly, \citet{xing2023quick} found that each ¥1 of government
subsidy generated approximately ¥3.07 in consumer spending in Shaoxing, China, and they further showed that a digital coupon initiative substantially increased consumption, with over 85\% of the spending occurring within three weeks. Most of the spending was genuinely incremental, although benefits were concentrated among
higher-end merchants, highlighting the need for careful program design to ensure inclusiveness.

Building on these insights, \citet{chen2025revenue} evaluated a program in Ningbo, China, and found that digital coupons significantly raised restaurant revenues, with a return of 4.5 yuan per 1 yuan of government spending. However, the effects were temporary, and while large businesses captured most absolute gains, smaller businesses experienced greater relative growth, contributing to reduced revenue and welfare inequality. These findings suggest that digital coupons are effective short-term support tools but are better suited for temporary relief rather than structural economic reform.

\section{Background and Data}

Taipei City is a small international metropolis and the primary
commercial center of Taiwan, with a population of approximately 2.7
million and a land area of about 271.8 square kilometers---equivalent to
only 38\% of Singapore's territory. In 2021, Taipei's GDP reached
approximately USD 233.5 billion, and in 2023, the average household
disposable income was around USD 35,000 \citep{DGBAS2024a,DGBAS2024b}.

Between 2021 and 2022, Taipei's economy suffered from the impacts of
COVID-19. Due to the near-complete cessation of domestic and
international tourism, industries such as hospitality, food and beverage
services, and retail experienced significant downturns. To revitalize
these sectors, the Taipei City Government launched two rounds of urban
digital consumption voucher programs: Taipei Bear Vouchers 1.0 in 2021
and Taipei Bear Vouchers 2.0 in 2022.

The Taipei Bear Vouchers 1.0 program issued five types of digital
consumption vouchers---accommodation, dining, cultural, sports, and
market vouchers---with a total of 5.457 million vouchers distributed and
a program budget of approximately NT\$ 553 million (around US\$17.2
million). Building on this foundation, the Taipei Bear Vouchers 2.0
program introduced an additional category, agricultural vouchers,
resulting in the issuance of approximately 690,000 vouchers and a
program budget of about NT\$ 584 million (around US\$18.25 million).
Compared to paper-based vouchers, digital consumption vouchers offer
several advantages. They not only reduce printing and administrative
costs but also enable rapid distribution and redemption via mobile
devices, effectively minimizing physical contact and helping to reduce
the risk of virus transmission during the pandemic.

Specifically, each accommodation voucher set comprised two NT\$500 vouchers, redeemable at participating hotels. Each dining voucher set included five NT\$100 vouchers, applicable at designated retail and food and beverage establishments, excluding chain convenience stores, supermarkets, e-commerce platforms, and entertainment venues. Each cultural, sports, and agricultural voucher set contained five NT\$100 vouchers, respectively usable at arts and cultural venues, public or private sports facilities, and farmers' markets. The market voucher set consisted of ten NT\$100 vouchers, which could be used at public markets, vendor-concentrated areas, and underground streets. Under the Taipei Bear Vouchers 2.0 program, each participant could register for up to two types of vouchers, which were randomly allocated. All vouchers were restricted for use within Taipei City and were not eligible for change, storage, resale, or transfer. To accelerate spending and stimulate urban consumption, the program set voucher expiration periods ranging from 45 to 60 days, aiming to encourage prompt redemption and support economic recovery.

In addition, the Taipei City Government introduced an extra round of the voucher program on December 16 of the same year. All individuals who had previously registered for the Taipei Bear Vouchers 2.0 lottery were eligible to receive additional vouchers corresponding to the types they originally selected, regardless of whether they had won in the first round. The values of these extra vouchers varied by type: NT\$500 for accommodation, and NT\$100 each for dining, cultural, sports, market, and agricultural vouchers. For instance, a participant who had registered for both accommodation and dining vouchers would receive one additional voucher of each type, totaling NT\$600 in value.

The dataset employed in this paper is derived from a survey conducted
among individuals who actually utilized the Taipei Bear Vouchers 2.0.
The survey was administered between March 1 and March 8, 2023, yielding
a total of 159,211 valid responses. Based on the survey results, this
paper estimates consumer behavioral parameters for each type of digital
voucher and are detailed in the following section.\footnote{To enhance the response rate, we offered a NT\$50 digital coupon (Family Mart) as a lottery incentive, which, according to prior studies (e.g., \citet{bosnjak2003prepaid}; \citet{sauermann2013increasing}), does not compromise response quality when used at a moderate level.}

\section{Empirical Method}
This section presents the empirical definitions and estimation methods
for the three principal quantities at the core of our analysis: the
expenditure substitution rate, the induced consumption rate, and the
impact of different levels of treatment intensity. We first describe how
survey responses facilitate the identification of these measures across
various voucher types. We then introduce a strategy for statistical
inference that explicitly incorporates the possibility of self-reporting
bias present in the survey data. Throughout the paper, we use
\(k = 1,2,\ldots,6\) to denote the six distinct types of vouchers issued
to participants, \(j = 1,...,J\) to index the mutually exclusive
subgroups \(g_{jk}\) within the respondent pool, each with
corresponding sample sizes \(n_{j,k}\), and define
\(n_{k} = n_{1,k} + ...n_{J,k}\) as the total number of respondents who
received voucher type \(k\).

\subsection{Expenditure Substitution}

The expenditure substitution rate is intended to capture the proportion
of voucher-financed consumption that would have occurred in the absence
of the voucher, that is, consumption that was already planned and for
which the voucher simply substituted for cash. To quantify this, each
respondent was asked the following question for each voucher type used:

\emph{``Did you make this consumption because you received type} \(k\)
\emph{voucher?''}

If a respondent answers ``Yes'', this indicates that the consumption was
directly triggered by the voucher, representing newly generated
spending. Conversely, a response of ``No'' signifies that the
consumption would have occurred regardless, and the voucher simply
replaced another method of payment, reflecting a substitution effect.
Based on these distinctions, we define the expenditure substitution rate
for subgroup \(j\) receiving voucher type \(k\) as follows:

\begin{align}
ES_{jk} = \frac{\sum_{i = 1}^{n_{j,k}}v_{ki}}{n_{j,k}}, \label{eq:es_j}
\end{align}
where \(v_{ki} = 1\) if respondent \(i\) answered ``No'', and
\(v_{ki} = 0\) if they answered ``Yes''. Accordingly,
\(ES_{jk}\) falls within the interval \(\lbrack 0,1\rbrack\),
where \(ES_{jk} = 1\) denotes full substitution, meaning all
spending would have taken place irrespective of the voucher, and
\(ES_{jk} = 0\) represents complete inducement, indicating that
all spending was newly generated due to the voucher. The expenditure
substitution rate for the entire group of respondents who received
voucher type \(k\) is defined in the same way, and given by:

\begin{align}
ES_{k} = \frac{\sum_{i = 1}^{n_{k}}v_{ki}}{n_{k}}. \label{eq:es}
\end{align}

\subsection{Induced Consumption}

The induced consumption rate captures the extent to which voucher usage
led to additional, out-of-pocket spending beyond the voucher's face
value. This metric is important for understanding whether vouchers
stimulated greater total consumption. Respondents were asked:

\emph{``When using the type $k$ voucher, did you incur any
additional spending beyond the value of type} \(k\) \emph{voucher?''}

Responses to this question were categorized into discrete spending
intervals, depending on the type of voucher received. The details of
these brackets are provided in the Appendix. Using these intervals,
the induced consumption rate for subgroup \(j\) receiving voucher
type \(k\) is computed as:
\begin{align}
IC_{jk} = \sum_{c = 1}^{C} \frac{m_{kc} \times b_{jkc,\mathrm{orig}}}{F_{k}^{\mathrm{orig}}} + \sum_{c = 1}^{C} \frac{m_{kc} \times b_{jkc,\mathrm{bonus}}}{F_{k}^{\mathrm{bonus}}}, \label{eq:ic_j}
\end{align}
where \(C\) denotes the total number of expenditure intervals, and
\(F_{k}^{l}\) represents the face value of voucher type \(k\) from
either the original or bonus round, with \(l \in \{\mathrm{orig}, \mathrm{bonus}\}\). The term \(m_{kc}\)
denotes the midpoint of interval \(c\); we follow the convention that
\(m_{kc} = 0\) for the interval indicating no additional spending,
and for the highest interval, \(m_{kc}\) is set to the lower bound of that uppermost range. We define:
\begin{align*}
b_{jkc,l} = \frac{1}{n_{j,k}} \sum_{i = 1}^{n_{j,k}} \mathbf{1}_{AS_i \in c \cap F_{ki} = F_k^{l}},
\end{align*}
where \(AS_i\) refers to the out-of-pocket additional spending reported by individual \(i\) in subgroup \(j\); \(F_{ki}\) denotes the face value of voucher type \(k\) received by individual \(i\), which can be either \(F_k^{\mathrm{orig}}\) or \(F_k^{\mathrm{bonus}}\), depending on the round of issuance; and the indicator function \(\mathbf{1}_{AS_i \in c \cap F_{ki} = F_k^l}\) equals 1 if \(AS_i\) falls in interval \(c\) and the face value of the voucher received by individual \(i\) is \(F_k^l\), and 0 otherwise.

Thus, the product \(m_{kc} \times b_{jkc,l}\) represents the average
additional spending within interval \(c\), and dividing by
\(F_{k}^{l}\) scales this amount relative to the corresponding
voucher face value. A value of \(IC_{jk} > 1\) implies that, on
average, respondents in subgroup \(j\) spent more out-of-pocket than
the voucher amount, suggesting a strong induced consumption effect.
Similarly, we define the overall induced consumption rate among all
respondents who received voucher type \(k\) as:
\begin{align}
IC_{k} = \sum_{c = 1}^{C} \frac{m_{kc} \times b_{kc,\mathrm{orig}}}{F_{k}^{\mathrm{orig}}} + \sum_{c = 1}^{C} \frac{m_{kc} \times b_{kc,\mathrm{bonus}}}{F_{k}^{\mathrm{bonus}}}, \label{eq:ic}
\end{align}
where
\[
b_{kc,l} = \frac{1}{n_k} \sum_{i = 1}^{n_k} \mathbf{1}_{AS_i \in c \cap F_{ki} = F_k^{l}}.
\]

\subsection{Intensity of Treatment}

Variation in treatment intensity arises from the presence of extra
stimulus vouchers issued alongside the original vouchers. These different
voucher schemes induce variation in the magnitude of economic stimulus
experienced by recipients. To measure the effect of such treatment
intensity, we define an index comparing average induced expenditures
between two groups: one that received the formal vouchers, and one that
received the additional stimulus vouchers without the original one. The intensity of treatment
effect for voucher \(k\) is defined as:

\begin{align}
IT_{k} = \sum_{c = 1}^{C}m_{kc} \times (b_{kc,\mathrm{orig}} - b_{kc,\mathrm{bonus}}),\label{eq:it}
\end{align}
where \(b_{kc,\mathrm{orig}}\) and \(b_{kc,\mathrm{bonus}}\) are the proportions of
respondents using original and additional stimulus vouchers whose reported
additional spending fall into interval \(c\), respectively. The difference
\(b_{kc,\mathrm{orig}} - b_{kc,\mathrm{bonus}}\) reflects the marginal effect of
moving from lower- to higher-intensity treatment for each spending
bracket, and it delivers different meanings compared with the induced
income rate.

\subsection{Identification and Inference}

This section sets out our identification strategy for estimating the
behavioral effects of voucher usage while accounting for potential
self-reporting bias in survey responses. The central analytical
challenge stems from the possibility that the reported outcome for
individual \(i\), who received voucher type \(k\), may not accurately
reflect the true effect owing to systematic distortions in reporting. To
formally model this concern, we adopt the following specification:

\begin{align}
y_{ik} = \theta_{ik} + B_{ik} + \epsilon_{ik},\notag
\end{align}
where \(y_{ik}\) represents the observed survey response (for
example, expenditure substitution \(v_{ki}\) or induced spending
\(m_{kc}\mathbf{1}_{AS_{i} \in c}/F_{k}\)),
\(\theta_{ik}\) denotes the unobserved true effect of the
voucher, \(B_{ik}\) captures the self-reporting bias, and
\(\epsilon_{ik}\) is a random error term with mean zero. Our
objective is to identify the average treatment effect,
\(\mathbb{E(}\theta_{ik})\). However, because \(y_{ik}\)
may be confounded by \(B_{ik}\), direct estimation using the
reported outcome does not isolate this quantity. To overcome this
limitation, we impose the following assumptions:

\textbf{Assumption 1~(Null Effects for Non-recipients)}. Individuals who do not receive any form of voucher are assumed to exhibit no behavioral response; specifically, both substitution and induced effects are zero for this group.

\textbf{Assumption 2~(Decomposition of True Effects)}.
The true treatment effect for individual \( i \) receiving voucher type \( k \), denoted \( \theta_{ik} \), can be decomposed as
\[
\theta_{ik} = \widetilde{\theta}_{ik} + \eta_{g_k(i)},
\]
where \( \widetilde{\theta}_{ik} \) captures the individual-specific component and \( \eta_{g_k(i)} \) reflects the deviation associated with subgroup \( g_k(i) = g_{jk} \). It is assumed that
\[
\mathbb{E}[\widetilde{\theta}_{ik}] = \theta_k \geq 0 \quad \text{and} \quad \mathbb{E}[\eta_{g_k(i)}] = 0.
\]

\textbf{Assumption 3~(Decomposition of Reporting Bias)}.
The self-reporting bias \( B_{ik} \) is similarly decomposed as
\[
B_{ik} = \widetilde{B}_{ik} + \nu_{g_k(i)},
\]
where \( \widetilde{B}_{ik} \) denotes the individual-level bias component and \( \nu_{g_k(i)} \) captures the group-specific deviation. We assume
\[
\mathbb{E}[\widetilde{B}_{ik}] = B_k \quad \text{and} \quad \mathbb{E}[\nu_{g_k(i)}] = 0.
\]
Additionally, we impose the constraint
\[
B_{ik} \times D \geq 0,
\]
where \( D \in \{-1, 1\} \) is a known sign indicator ensuring that the reporting bias is one-sided (e.g., either always nonnegative or nonpositive depending on the context).

The first assumption establishes a natural benchmark: if a respondent
did not receive any voucher, then by definition there is no channel
through which substitution or induced spending could occur. Hence,
\(\theta_{ik} = 0\) when individual \(i\) is not exposed to
voucher \(k\). This condition allows us to interpret positive reported
values as attributable to the treatment, and rules out baseline
confounding from untreated individuals. The second assumption allows for
heterogeneity in the true effect across subgroups, captured by the group
effect \(\eta_{g_{k}(i)} = \eta_{g_{jk}}\). However, this
heterogeneity averages out across groups, resulting in an overall
unconditional mean of \(\theta_{k}\). For instance, the effect of a
dining voucher might differ by age or region, and this structure permits
such variation without requiring full individual-level modeling of
unobserved heterogeneity. The third assumption mirrors the structure of
the second and allows the self-reporting bias to vary across groups
through \(\nu_{g_{k}(i)}\). These assumptions are consistent with
empirical findings, such as those reported by \citet{geisen2012examining},
which document systematic variation in survey reporting accuracy across
demographic groups.

Under these assumptions, the group-specific expected value of the
observed outcome satisfies:

\begin{align}
\mathbb{E(}y_{ik} \mid g_{k}(i) = g_{jk}) = \theta_{k} + \eta_{g_{k}(i)} + B_{k} + \nu_{g_{k}(i)} = \theta_{jk} + B_{jk},\notag
\end{align}
where \(\theta_{jk} = \theta_{k} + \eta_{g_{jk}}\) and
\(B_{jk} = B_{k} + \nu_{g_{jk}}\). Since
\(\theta_{jk}\) and \(B_{jk}\) are not separately
identified, we estimate a lower bound on \(\theta_{jk}\) using a
conservative bias correction. Specifically, we define:

\begin{align}
{\widetilde{y}}_{jk}^{\text{lower}} & = {\widehat{y}}_{jk} - {\widehat{B}}_{k}, \label{eq:lower} \\
{\widetilde{y}}_{jk}^{\text{upper}} & = {\widehat{y}}_{jk}, \label{eq:upper}
\end{align}
where \({\widehat{y}}_{jk}\) could be either \(ES_{jk}\)
or \(IC_{jk}\), and
\({\widehat{B}}_{k} = \min_{j}{\widehat{y}}_{jk}\) represents the
most conservative estimate of the bias. On this basis, we derive an
interval estimate for the true effect \(\theta_{jk}\), where the
lower bound is adjusted to account for potential bias, and the upper
bound corresponds to the unadjusted sample mean. This correction strategy does not assert that the true effect in the lowest observed subgroup is literally zero. Rather, it adopts a worst-case bounding perspective, treating the minimum observed outcome as an upper bound on potential bias. That is,

\begin{align}
B_{jk} \leq \min_{j}{\widehat{y}}_{jk},\quad\forall j.\notag
\end{align}
Furthermore, the estimated bias can be applied to either \(ES_{k}\) or
\(IC_{k}\) when constructing lower and upper bounds, whether for the
entire sample or for broader groups formed by aggregating finer
subgroups \(j\)s into their respective parent groups.

To quantify the uncertainty of these estimates, we implement a
stratified bootstrap procedure to construct \(100(1 - \alpha)\%\)
confidence intervals for both bounds as shown in Algorithm 1.\footnote{Formally, the minimum operator is not fully differentiable only when multiple subgroup estimates tie exactly. Given that self-reporting bias is expected to vary across groups, such ties occur with probability zero asymptotically. Hence, the condition in \citet[Thm 3.1]{fang2019inference} for bootstrap consistency is satisfied.}

\subsection{Regional Input--Output Analysis}

The implementation of consumption voucher programs can stimulate private
consumption expenditures, thereby increasing overall final demand within
the economic system. To meet the additional demand, industries expand
the supply of goods and services, further invigorating economic
activities.

In the case of Taipei's digital consumption voucher programs, each type
of voucher was restricted for use within specific industries, implying
that the initial stimulus effects would be concentrated in the outputs
of those targeted sectors. As these sectors expand their output,
inter-industry linkages subsequently induce additional output growth
across other sectors. This process can be represented as follows:
\begin{align}
\MBy =  \left(\MBI - \MBA \right)^{- 1} \times \left( \mathrm{\Delta}\MBF \right) \circ VA, \label{eq:io}
\end{align}
where $\MBy$ is a vector representing 19 industrial sectors' output in Taipei, $\MBA$ denotes the corresponded input-output coefficients calculated based on the input-output table released in 2016, \( \mathrm{\Delta}\MBF\)
represents the changes in final demand across industries, and $VA$ denotes the sector-specific value-added coefficients. Based on Equation \eqref{eq:io}, we can further calculate the cumulative effect from a change in final demand induced by the consumption voucher
program propagates through inter-industry linkages, leading to
corresponding changes in industrial output.

In addition, since Taiwan's official statistics provide input–output tables only at the national level, this paper constructs a Taipei City input–output table by combining the national input–output table with regional statistics from Taiwan's Industry and Service Census Report, including data on output, value added. This regional table serves as the basis for subsequent analysis. For details on the methodology and technical procedures used to compile the Taipei City input–output table, please refer to the Appendix.

\begin{algorithm}[H]
\caption{Stratified Bootstrap for Bounds}
\begin{algorithmic}[1]
\REQUIRE Original data $\{y_{ik}\}$, groups $g_{1k},\dots,g_{Jk}$, replications $B_s$, significance $\alpha$
\ENSURE Bootstrap CIs $\mathrm{CI}_{jk}^{\mathrm{lower}}$, $\mathrm{CI}_{jk}^{\mathrm{upper}}$
\FOR{$r = 1$ \TO $B_s$}
  \FOR{each group $j=1,\dots,J$}
    \STATE Resample $n_{j,k}$ observations $\{y_{ik}^*\}_{i\in g_{jk}}$ with replacement
    \STATE Compute $\hat{y}_{jk}^{*(r)}$
    \STATE Set $b_k^{*(r)} = \min_j \hat{y}_{jk}^{*(r)}$
  \ENDFOR
  \STATE For each $j$, define:
    \[
    \tilde{y}_{jk}^{\mathrm{lower},*(r)} = \hat{y}_{jk}^{*(r)} - B_k^{*(r)}, \quad
    \tilde{y}_{jk}^{\mathrm{upper},*(r)} = \hat{y}_{jk}^{*(r)}.
    \]
  \STATE Store $\tilde{y}_{jk}^{\mathrm{lower},*(r)}$, $\tilde{y}_{jk}^{\mathrm{upper},*(r)}$
\ENDFOR
\FOR{each group $j=1,\dots,J$}
  \STATE Compute:
    \begin{align}
    \mathrm{CI}_{jk}^{\mathrm{lower}} = \left[
      Q_{\alpha/2}\left(\{\tilde{y}_{jk}^{\mathrm{lower},*(r)}\}\right), \,
      Q_{1-\alpha/2}\left(\{\tilde{y}_{jk}^{\mathrm{lower},*(r)}\}\right)
    \right]\label{eq:lowerCI}
    \end{align}
  \STATE Compute:
    \begin{align}
    \mathrm{CI}_{jk}^{\mathrm{upper}} = \left[
      Q_{\alpha/2}\left(\{\tilde{y}_{jk}^{\mathrm{upper},*(r)}\}\right), \,
      Q_{1-\alpha/2}\left(\{\tilde{y}_{jk}^{\mathrm{upper},*(r)}\}\right)
    \right]\label{eq:upperCI}
    \end{align}
\ENDFOR
\RETURN $\mathrm{CI}_{jk}^{\mathrm{lower}}$, $\mathrm{CI}_{jk}^{\mathrm{upper}}$
\end{algorithmic}
\end{algorithm}

\section{Empirical Results}

\subsection{Summary Statistics}

Table \ref{Tab:tab1} demonstrates the sample structure of the Taipei Bear Vouchers
2.0 user survey. A total of 159,211 valid responses were collected.
Among the voucher types, dining vouchers accounted for the highest share
at 64.5\%, followed by market vouchers and cultural vouchers, with
shares of 22.7\% and 6.4\%, respectively. Overall, the relative
proportions of the survey samples for different voucher types were
consistent with the relative proportions of actual policy expenditures.

\begin{table}
\begin{footnotesize}
\begin{center}
\caption{Sample Structure of the Taipei City Digital Consumption Voucher
Survey}
\label{Tab:tab1}
\begin{tabular}{lcccccccc}
\toprule
& \multicolumn{8}{c}{Number of Valid Samples}\\
\cmidrule(l){2-9}
& \multicolumn{2}{c}{Gender} & \multicolumn{2}{c}{Residence} & \multicolumn{3}{c}{Age} & \multicolumn{1}{c}{Overall} \\
\cmidrule(l){2-3}\cmidrule(l){4-5}\cmidrule(l){6-8}
& Male & Female & Taipei & Other Cities & $<30$ & $30-49$ & $>49$ & \\
\midrule
Accommodation Voucher & 699 & 932 & 626 & 1,005 & 439 & 919 & 273 &
1,631 \\
Dining Voucher & 36,962 & 65,773 & 51,643 & 51,092 & 16,917 & 52,108 &
33,710 & 102,735 \\
Cultural Voucher & 3,234 & 6,844 & 4,971 & 5,107 & 2,655 & 5,430 &
1,993 & 10,078 \\
Sports Voucher & 1,204 & 1,814 & 2,069 & 949 & 645 & 1,579 & 794 &
3,018 \\
Market Voucher & 13,572 & 22,616 & 18,259 & 17,929 & 4,963 & 17,242 &
13,983 & 36,188 \\
Agricultural Voucher & 1,791 & 3,780 & 3,095 & 2,476 & 436 & 2,423 &
2,712 & 5,571 \\
Total & 57,462 & 101,759 & 80,663 & 78,558 & 26,055 & 79,701 & 53,465 &
159,221 \\
\midrule
& \multicolumn{8}{c}{Percentage of Valid Samples (\%)}\\
\cmidrule(l){2-9}
Accommodation Voucher & 42.9 & 57.1 & 38.4 & 61.6 & 26.9 & 56.3 & 16.7
& 1.0 \\
Dining Voucher & 36.0 & 64.0 & 50.3 & 49.7 & 16.5 & 50.7 & 32.8 &
64.5 \\
Cultural Voucher & 32.1 & 67.9 & 49.3 & 50.7 & 26.3 & 53.9 & 19.8 &
6.4 \\
Sports Voucher & 39.9 & 60.1 & 68.6 & 31.4 & 21.4 & 52.3 & 26.3 &
1.9 \\
Market Voucher & 37.5 & 62.5 & 50.5 & 49.5 & 13.7 & 47.6 & 38.6 &
22.7 \\
Agricultural Voucher & 32.1 & 67.9 & 55.6 & 44.4 & 7.8 & 43.5 & 48.7 &
3.5 \\
Total & 36.1 & 63.9 & 50.7 & 49.3 & 16.4 & 50.1 & 33.6 & 100.0 \\
\bottomrule
\end{tabular}
\parbox{15cm}{
\vspace{0.2cm}
Note: The survey was conducted via the TaipeiPASS platform and targeted consumers who had both been selected to receive and had used a specific type of consumption voucher. The survey period spanned from March 1 to
March 8, 2023.}
\end{center}
\end{footnotesize}
\end{table}

Regarding gender, since the TaipeiPASS platform does not require users
to provide gender information during registration, the actual gender
distribution of voucher users remains unknown. However, based on the
survey, female respondents accounted for 63.9\%, suggesting that the
proportion of female users may be higher than that of male users. The
distribution of respondents was roughly evenly split between residents
of Taipei City and those from other cities. Notably, the proportion of
accommodation voucher users from other cities was higher than that from
Taipei City, implying a greater demand among non-Taipei residents for
traveling to Taipei City to utilize the vouchers for accommodation
purposes. Finally, with respect to age, respondents under the age of 50
accounted for a higher proportion of responses, reflecting a relatively
lower level of digital tool usage among older consumers compared to
younger age groups.

\subsection{Expenditure Substitution Effect}
As previously discussed, the substitution effect plays a crucial role in determining the effectiveness of consumption voucher policies. Holding other factors constant, a higher rate of expenditure substitution implies a lower net economic benefit arising from the stimulus vouchers.

\begin{figure}[htbp]
    \centering
    \begin{minipage}{0.45\textwidth}
        \centering
        \includegraphics[width=\linewidth]{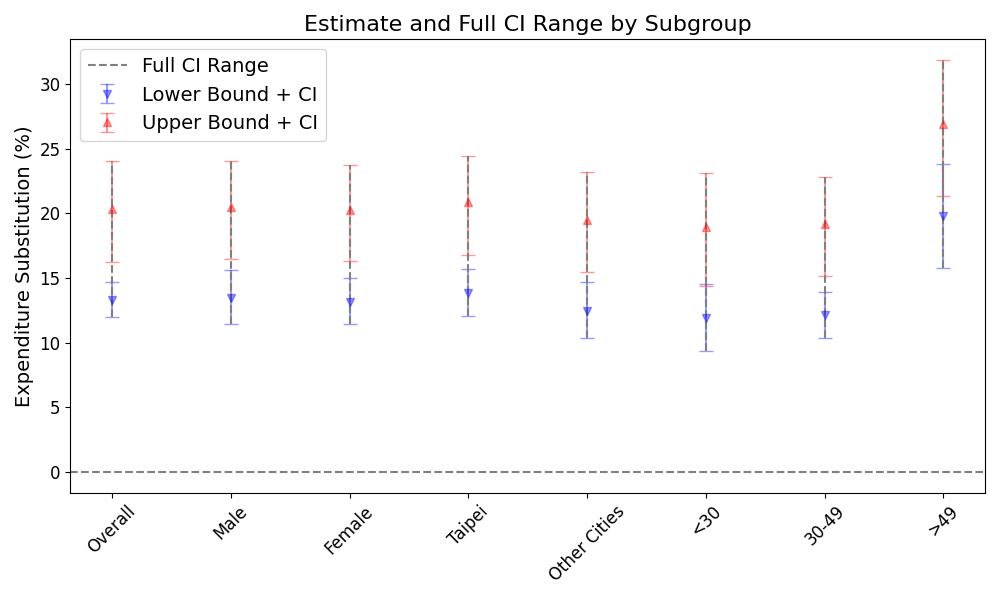}
        \subcaption{Accommodation}
    \end{minipage}
    \hfill
    \begin{minipage}{0.45\textwidth}
        \centering
        \includegraphics[width=\linewidth]{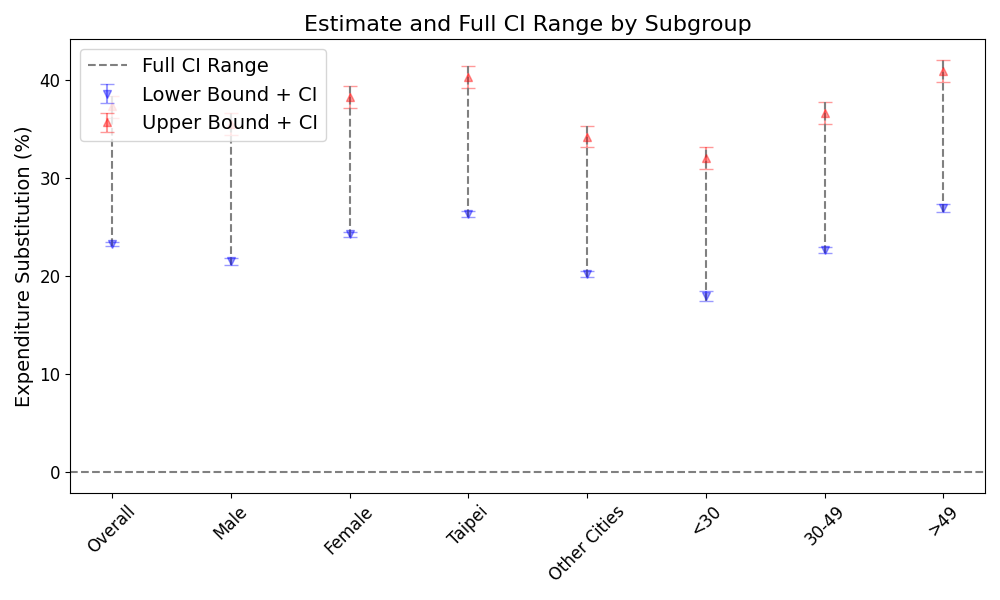}
        \subcaption{Dining}
    \end{minipage}
    \vspace{1em}
    \begin{minipage}{0.45\textwidth}
        \centering
        \includegraphics[width=\linewidth]{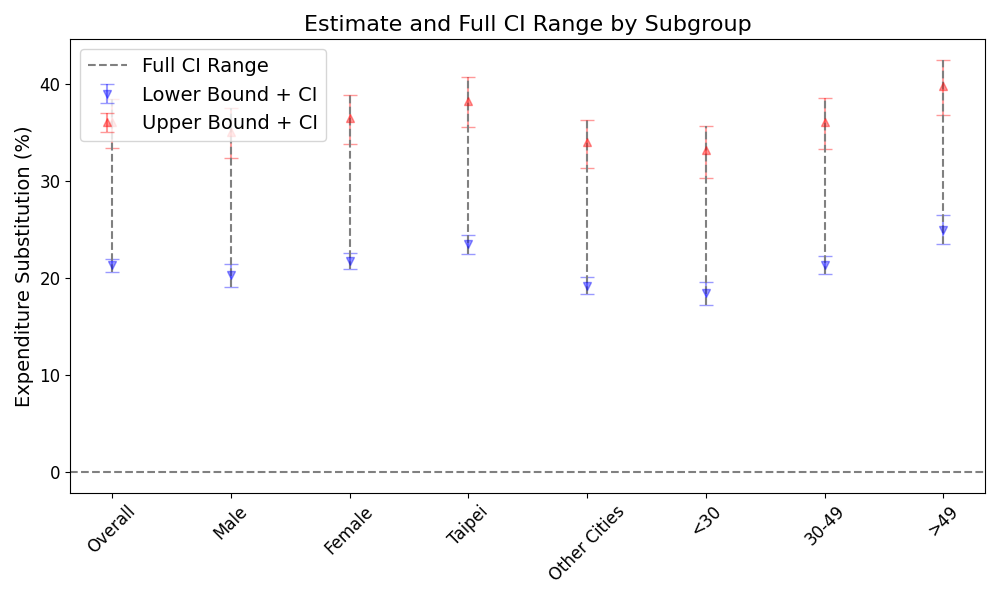}
        \subcaption{Cultural}
    \end{minipage}
    \hfill
    \begin{minipage}{0.45\textwidth}
        \centering
        \includegraphics[width=\linewidth]{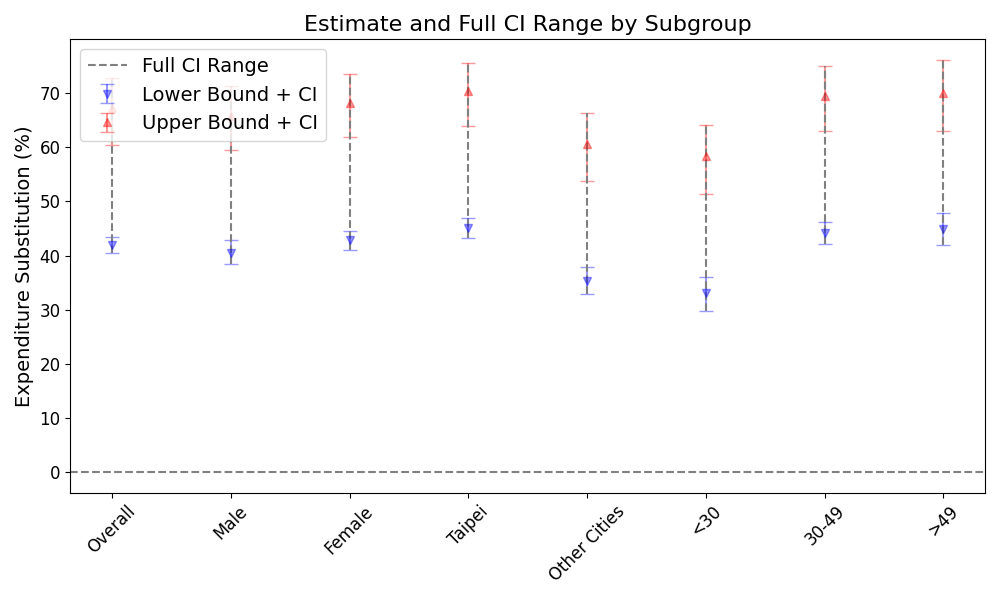}
        \subcaption{Sports}
    \end{minipage}
    \vspace{1em}
    \begin{minipage}{0.45\textwidth}
        \centering
        \includegraphics[width=\linewidth]{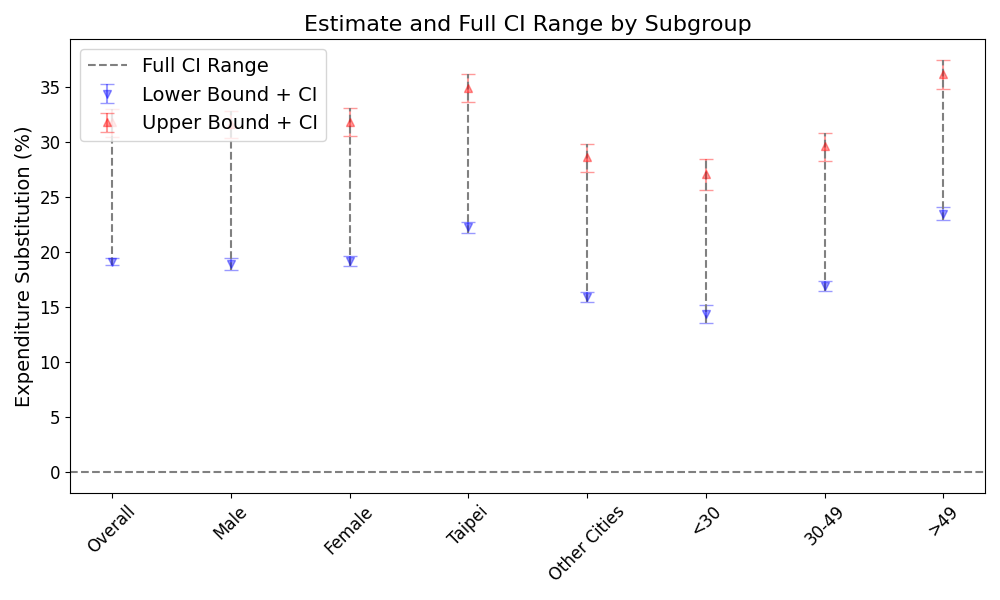}
        \subcaption{Market}
    \end{minipage}
    \hfill
    \begin{minipage}{0.45\textwidth}
        \centering
        \includegraphics[width=\linewidth]{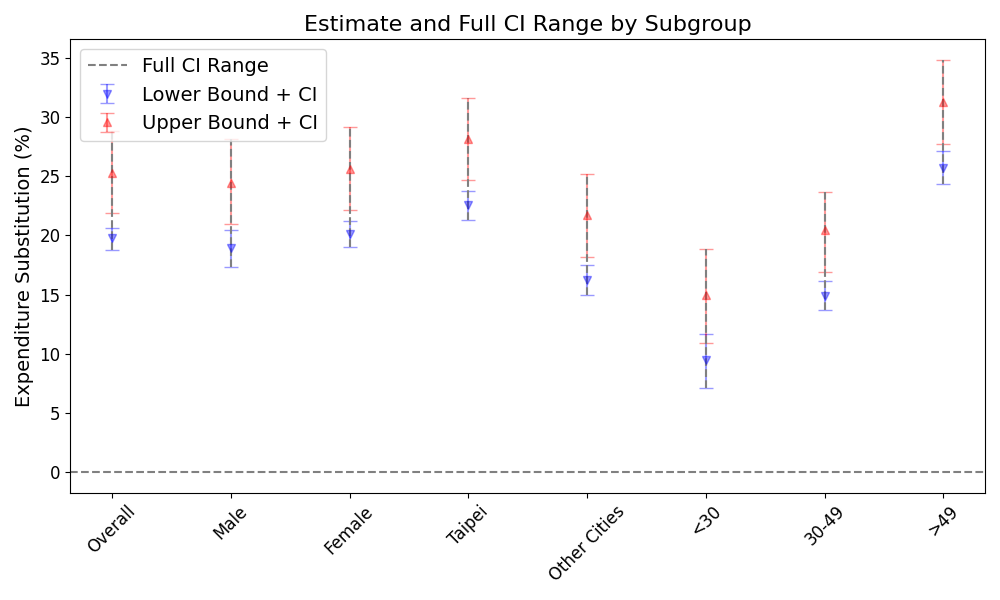}
        \subcaption{Agricultural}
    \end{minipage}
    \caption{Expenditure Substitution Rate}
    \label{fig:se}
\end{figure}

The main empirical results are presented in Table~\ref{Tab:tab2} and visualized in Figure~\ref{fig:se}. Table~\ref{Tab:tab2} reports substitution rates for six voucher types across subgroups defined by gender, residence, and age, along with the overall substitution rate shown in the final column. For each combination of voucher type and subgroup, we report the lower-est and upper-est values implied by the confidence region obtained via the stratified bootstrap method described in Section~4.4. The lower-est value and upper-est value represent optimistic and pessimistic bounds on the substitution rate, respectively.

Among all types, the sports voucher exhibits the highest substitution rate, ranging from 40.5\% to 72.8\%, suggesting that sports-related expenditures exhibit relatively inelastic demand. This likely reflects that participation in sports is a habitual behavior, with most voucher users already engaged in regular exercise. Furthermore, restrictions limiting the sports voucher to use at sports facilities or related merchandise reinforce the observed substitution pattern. In contrast, the accommodation voucher exhibits a much lower substitution rate, ranging from 12.0\% to 24.0\%, indicating that the voucher is more likely to generate new consumption rather than substitute pre-planned spending. Dining, cultural, market, and agricultural vouchers fall within a moderate range of 18.8\% to 38.5\%.

Compared to the substitution rates observed under the Taipei Bear Vouchers 1.0 program \citep{taipei2022bear}, the accommodation voucher also recorded the lowest rate (approximately 8\%), while the sports voucher exhibited the highest (around 39\%). The substitution rates for the dining, market, and cultural vouchers ranged between 24\% and 30\%, indicating a degree of behavioral inertia among consumers in their use of consumption vouchers. Notably, the substitution rates for the accommodation and sports vouchers were lower in the Taipei Bear Vouchers 1.0 program than in Taipei Bear Vouchers 2.0, whereas the other voucher types displayed comparatively lower substitution rates in the Taipei Bear Vouchers 2.0 program.\footnote{\citet{taipei2022bear} does not report substitution rate data for the Taipei Bear Vouchers 1.0 program disaggregated by demographic characteristics.}

Figure~\ref{fig:se} further illustrates the confidence intervals for the estimated substitution rates. Notably, individuals aged above 49 consistently exhibit higher substitution rates across all voucher types, suggesting that older adults tend to follow more planned consumption patterns. For dining, cultural, and agricultural vouchers, the substitution rates show an upward trend with age. These differences are statistically significant when considering the lower bounds and their associated confidence intervals, and remain observable when examining the upper bounds, particularly if reporting bias is taken into account.

Regarding residence, the substitution rate is higher for individuals residing in Taipei across all voucher types, indicating a pronounced tendency to use vouchers to replace planned consumption. This finding aligns with the consumption behavior heterogeneity highlighted by \citet{johnson2006household}, suggesting that residents of Taipei and other urban areas may differ in socioeconomic characteristics that influence their marginal propensity to consume. In contrast, gender differences appear to have no statistically significant effect on substitution rates according to our survey results.

\begin{landscape}
\begin{table}
\begin{footnotesize}
\begin{center}
\caption{Expenditure Substitution Effect}
\label{Tab:tab2}
\begin{tabular}{lcccccccc}
\toprule
& \multicolumn{2}{c}{Gender} & \multicolumn{2}{c}{Residence} & \multicolumn{3}{c}{Age} & \multicolumn{1}{c}{Overall} \\
\cmidrule(l){2-3}\cmidrule(l){4-5}\cmidrule(l){6-8}
& Male & Female & Taipei & Other Cities & $<30$ & $30-49$ & $>49$ & \\
\midrule
Accommodation Voucher & [11.4\%, 24.1\%] & [11.5\%, 23.7\%] & [12.0\%, 24.4\%] & [10.4\%, 23.2\%] & [9.3\%, 23.1\%] & [10.3\%, 22.8\%] & [15.8\%, 31.9\%] & [12.0\%, 24.0\%] \\
Dining Voucher & [21.2\%, 36.6\%] & [24.0\%, 39.4\%] & [26.0\%, 41.4\%] & [19.9\%, 35.3\%] & [17.5\%, 33.1\%] & [22.3\%, 37.7\%] & [26.5\%, 42.0\%] & [23.0\%, 38.4\%] \\
Cultural Voucher & [19.1\%, 37.5\%] & [20.9\%, 38.9\%] & [22.5\%, 40.7\%] & [18.4\%, 36.3\%] & [17.2\%, 35.7\%] & [20.4\%, 38.5\%] & [23.5\%, 42.5\%] & [20.6\%, 38.5\%] \\
Sports Voucher & [38.4\%, 71.4\%] & [41.0\%, 73.5\%] & [43.2\%, 75.6\%] & [32.9\%, 66.2\%] & [29.8\%, 64.1\%] & [42.1\%, 74.9\%] & [41.9\%, 76.1\%] & [40.5\%, 72.8\%] \\
Market Voucher & [18.4\%, 32.8\%] & [18.7\%, 33.1\%] & [21.7\%, 36.2\%] & [15.5\%, 29.8\%] & [13.6\%, 28.4\%] & [16.4\%, 30.8\%] & [22.9\%, 37.4\%] & [18.8\%, 33.0\%] \\
Agricultural Voucher & [17.4\%, 28.2\%] & [19.0\%, 29.2\%] & [21.3\%, 31.6\%] & [15.0\%, 25.2\%] & [7.1\%, 18.8\%] & [13.7\%, 23.7\%] & [24.4\%, 34.8\%] & [18.8\%, 28.8\%] \\
\bottomrule
\end{tabular}
\parbox{24cm}{
\vspace{0.2cm}
Note: The substitution rate was measured based on respondents' answers
to the following question: \emph{"Did you make this consumption because
you received the \underline{specific voucher}?"} If the respondent
answered "No," it indicated that the consumption behavior was not
induced by the receipt of the voucher. That is, the voucher was used to
pay for planned expenditures rather than to generate additional
consumption. For detailed survey content, please refer to the appendix.
The method for calculating the substitution rate is presented in
Equations \eqref{eq:es_j} and \eqref{eq:es}.}
\end{center}
\end{footnotesize}
\end{table}
\end{landscape}

\subsection{Induced Consumption Effect}
The induced consumption effect refers to the additional spending generated by consumers when utilizing consumption vouchers. Holding other factors constant, a higher induced consumption effect indicates greater economic benefits resulting from the voucher program.

\begin{figure}[htbp]
    \centering
    \begin{minipage}{0.45\textwidth}
        \centering
        \includegraphics[width=\linewidth]{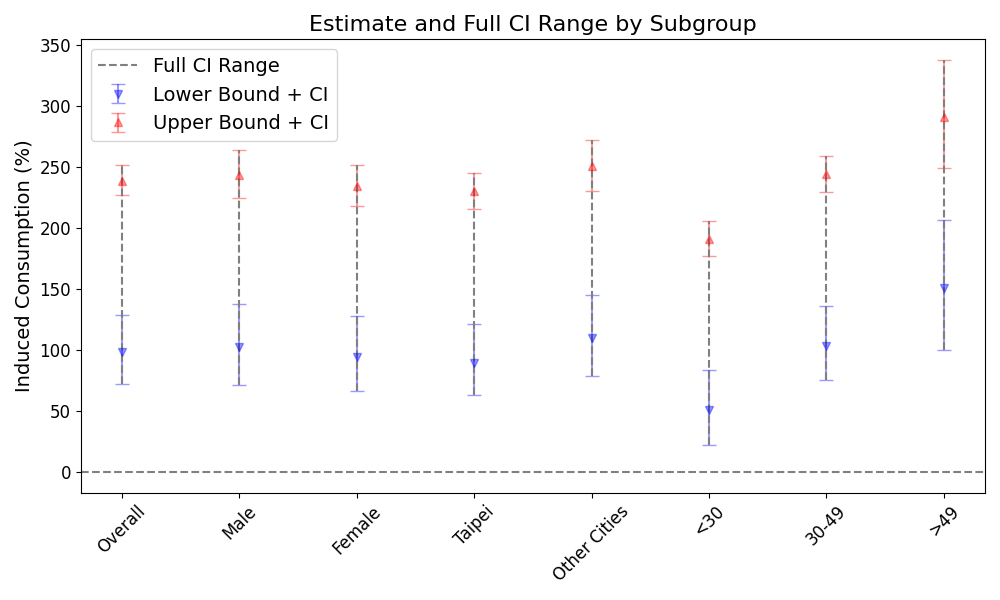}
        \subcaption{Accommodation}
    \end{minipage}
    \hfill
    \begin{minipage}{0.45\textwidth}
        \centering
        \includegraphics[width=\linewidth]{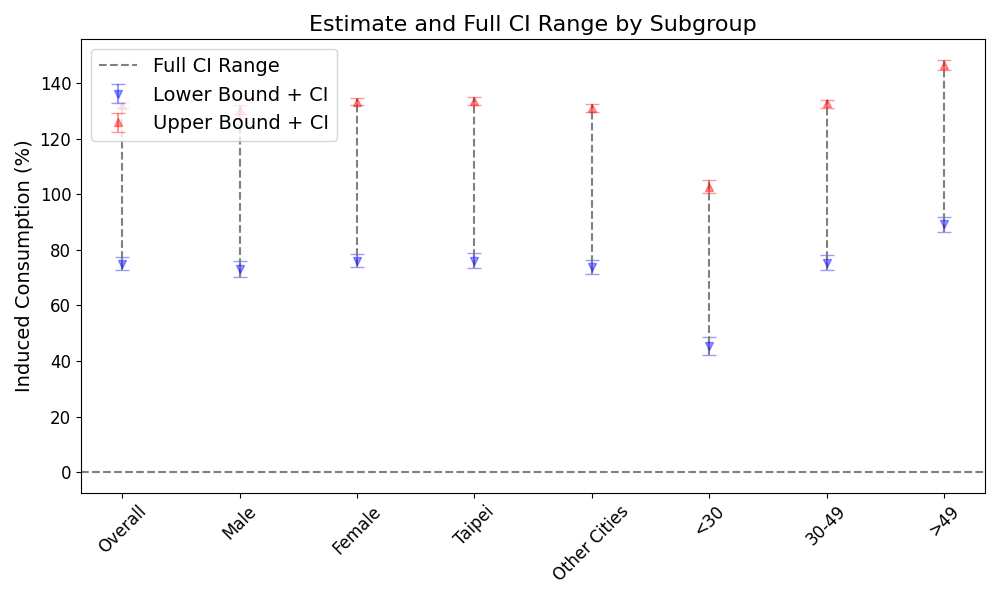}
        \subcaption{Dining}
    \end{minipage}
    \vspace{1em}
    \begin{minipage}{0.45\textwidth}
        \centering
        \includegraphics[width=\linewidth]{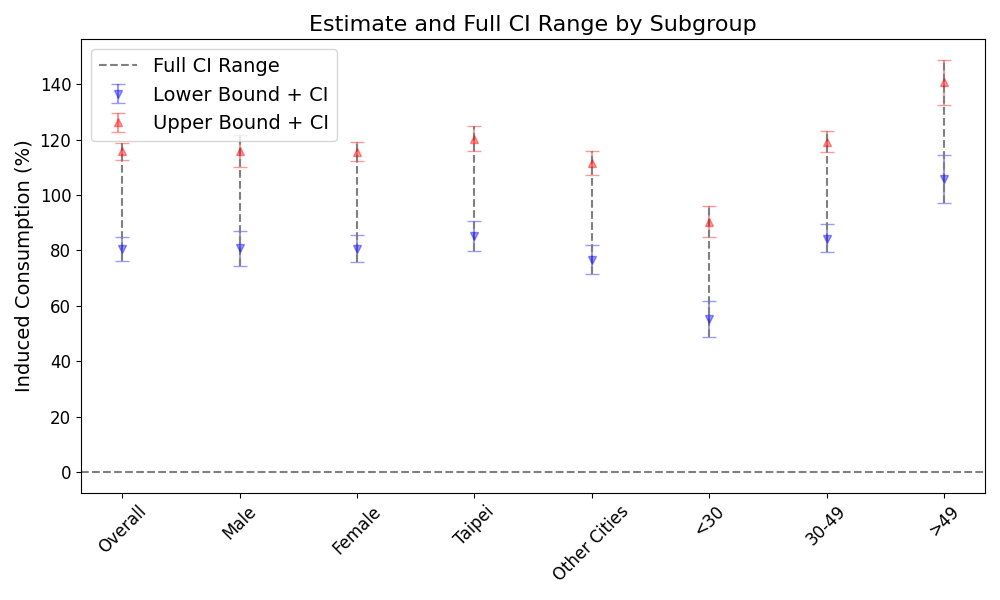}
        \subcaption{Cultural}
    \end{minipage}
    \hfill
    \begin{minipage}{0.45\textwidth}
        \centering
        \includegraphics[width=\linewidth]{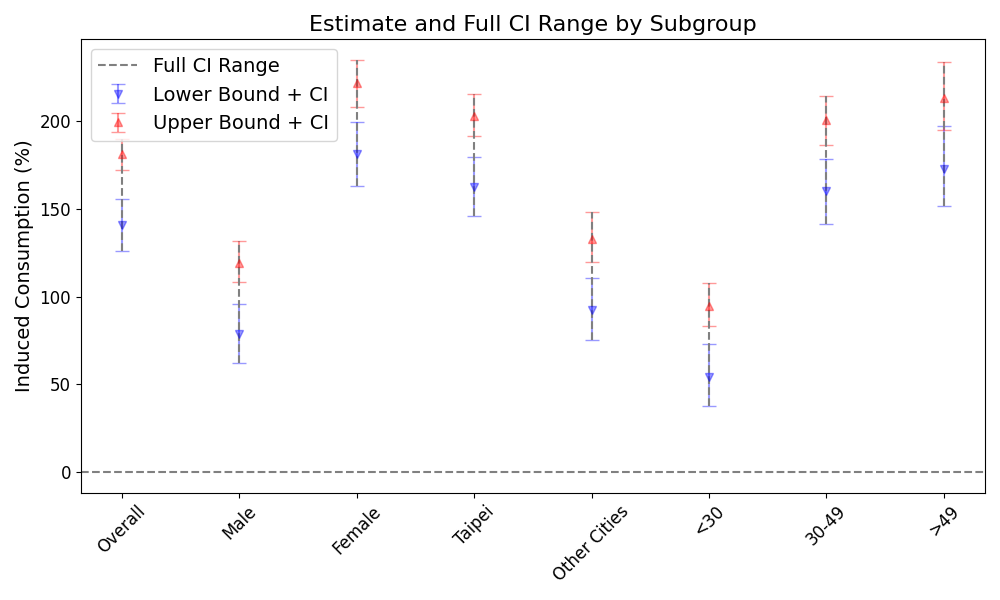}
        \subcaption{Sports}
    \end{minipage}
    \vspace{1em}
    \begin{minipage}{0.45\textwidth}
        \centering
        \includegraphics[width=\linewidth]{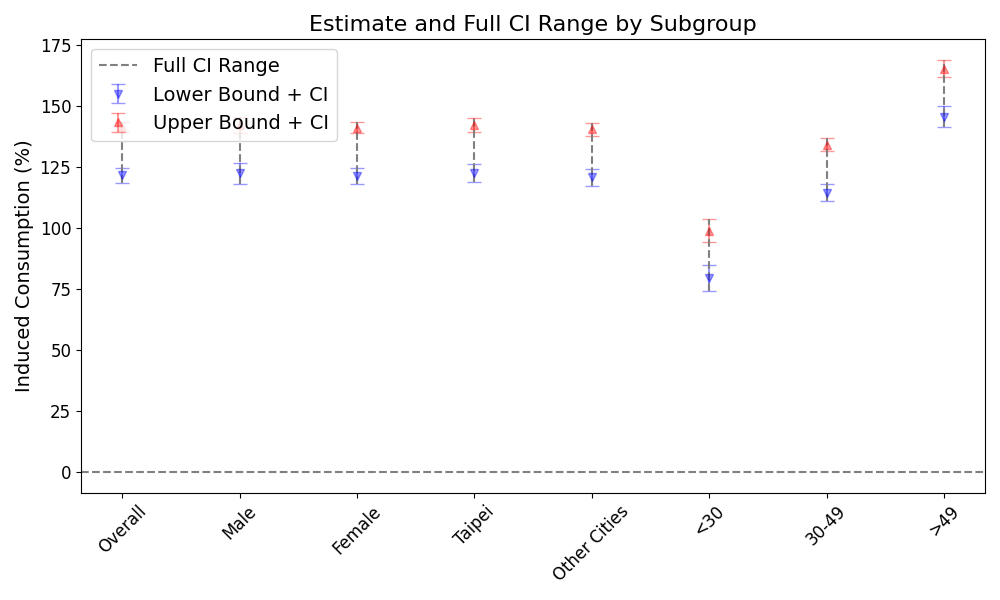}
        \subcaption{Market}
    \end{minipage}
    \hfill
    \begin{minipage}{0.45\textwidth}
        \centering
        \includegraphics[width=\linewidth]{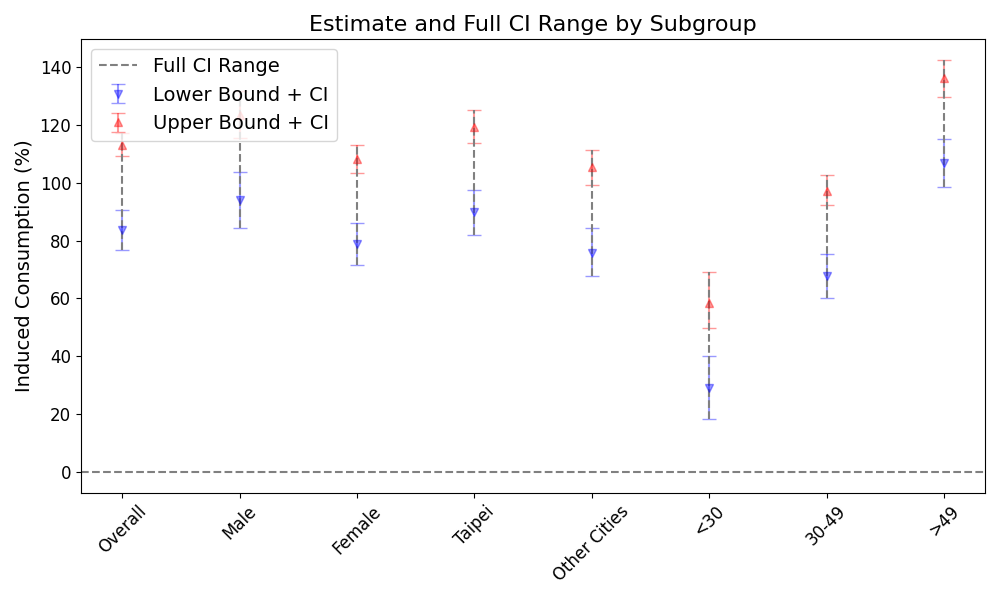}
        \subcaption{Agricultural}
    \end{minipage}
    \caption{Expenditure Substitution Rate}
    \label{fig:ic}
\end{figure}

The estimated induced consumption rates are reported in Table~\ref{Tab:tab3} and visualized in Figure~\ref{fig:ic}. Similar to the analysis of substitution effect, we report results across subgroups defined by demographic characteristics, with the overall outcome presented in the final column. Each cell contains both the lower-est and upper-est values, representing pessimistic and optimistic bounds, respectively. It is important to note that the definitions of pessimism and optimism here are the reverse of those in the substitution rate analysis.\footnote{This reversal arises because a higher induced consumption effect reflects a more substantial stimulus impact of the vouchers.}

As shown in Table~\ref{Tab:tab3}, the accommodation voucher yields the most pronounced induced consumption effect, with an overall range from 72.5\% to 251.6\%. This suggests that consumers typically spend an additional amount approximately twice the face value of the voucher when using the accommodation voucher. The sports voucher also demonstrates a strong induced consumption effect, despite being associated with the highest substitution rate. This may reflect the relatively high price level of sports-related goods and services, which implies that the baseline expenditure in this category is substantial. Furthermore, the market voucher generates a larger induced consumption effect relative to the dining, cultural, and agricultural vouchers.

As an additional note, the induced consumption effect observed under the Taipei Bear Vouchers 2.0 program differ notably from those recorded in the Taipei Bear Vouchers 1.0 program. According to the \citet{taipei2022bear}, under the 1.0 program, the sports voucher generated the highest induced consumption effect, approximately 100\%, followed by the dining voucher at around 79\%. In contrast, the market, cultural, and accommodation vouchers exhibited relatively lower induced consumption rates, ranging from 11\% to 40\%.\footnote{\citet{taipei2022bear} does not report induced consumption effect for the Taipei Bear Vouchers 1.0 program disaggregated by demographic characteristics.}

Turning to Figure~\ref{fig:ic}, a striking result is that the induced consumption effect increases monotonically across age groups for all voucher types. Even when focusing on the lower bounds and their confidence intervals (accounting for potential reporting bias), the effects remain statistically distinguishable across age groups for the dining, cultural, sports, market, and agricultural vouchers. This finding indicates that older consumers are more likely to engage in additional spending, even though they also tend to exhibit higher substitution rates.

Moreover, the sports voucher presents a particularly distinctive pattern. First, when examining gender-based subgroups, the induced consumption effect is substantially higher for females than for males. This finding is consistent with \citet{lee2024exploring} and related references, which suggest that female consumers are increasingly driven by self-motivated consumption factors nowadays. Second, a notably strong induced consumption rate is observed among respondents residing in Taipei, potentially reflecting higher levels of discretionary spending on premium sports-related goods and services.

\subsection{Intensity of Treatment Effect}
The issuance of extra stimulus vouchers alongside the original vouchers constitutes a natural experiment that facilitates the identification of induced consumption under varying face values. As previously noted, the values of these additional vouchers differ by type and are NT\$500, NT\$100, NT\$100, NT\$100, NT\$100, and NT\$100 for accommodation, dining, cultural, sports, market, and agricultural vouchers, respectively. Individuals who registered for the first-stage voucher lottery received two specific voucher types in the second round, regardless of whether they had won in the first stage. The face values of the first-stage vouchers were NT\$1,000, NT\$500, NT\$500, NT\$500, NT\$1,000, and NT\$500 for accommodation, dining, cultural, sports, market, and agricultural vouchers, respectively. This structure enables a comparison between respondents who received vouchers in the first round and those who only received them due to the second-stage bonus policy. Based on Equation~\eqref{eq:it} and the stratified bootstrap procedure, we summarize the estimated induced consumption effects across varying voucher intensities in Table~\ref{Tab:tab4}.

Unsurprisingly, the accommodation voucher exhibits the largest induced consumption effect, consistent with its higher additional face value of NT\$500=NT\$1,000-NT\$500. The results indicate that it generates a multiplier greater than one within the 95\% confidence interval. Among the vouchers with an additional value of NT\$400=NT\$500-NT\$100, the dining, cultural, and agricultural vouchers also lead to induced consumption increases exceeding 25\%. In contrast, the sports and market vouchers show relatively smaller effects, despite their additional face values of NT\$400 and NT\$900=NT\$1,000-NT\$100, respectively. While these figures may suggest a limited marginal impact from increasing voucher values, they also offer insight from two perspectives. First, since the second-round bonus vouchers were unexpected, recipients may have treated them as outside their planned expenditure, thus contributing to incremental consumption. Second, given that participating merchants had already implemented promotional campaigns aligned with the program, even a relatively small face value could effectively stimulate consumer spending.

\begin{landscape}
\begin{table}
\begin{footnotesize}
\begin{center}
\caption{Induced Consumption Effect}
\label{Tab:tab3}
\begin{tabular}{lcccccccc}
\toprule
& \multicolumn{2}{c}{Gender} & \multicolumn{2}{c}{Residence} & \multicolumn{3}{c}{Age} & \multicolumn{1}{c}{Overall} \\
\cmidrule(l){2-3}\cmidrule(l){4-5}\cmidrule(l){6-8}
& Male & Female & Taipei & Other Cities & $<30$ & $30-49$ & $>49$ & \\
\midrule
Accommodation Voucher & [71.4\%, 264.0\%] & [66.4\%, 252.1\%] & [62.9\%, 245.5\%] & [79.1\%, 272.6\%] & [22.4\%, 206.2\%] & [75.9\%, 259.3\%] & [100.1\%, 338.1\%] & [72.5\%, 251.6\%] \\
Dining Voucher & [70.2\%, 132.2\%] & [73.6\%, 134.5\%] & [73.5\%, 134.9\%] & [71.2\%, 132.5\%] & [42.1\%, 105.1\%] & [72.9\%, 134.1\%] & [86.2\%, 148.3\%] & [72.7\%, 133.3\%] \\
Cultural Voucher & [74.4\%, 121.7\%] & [76.0\%, 119.3\%] & [79.8\%, 124.8\%] & [71.3\%, 115.8\%] & [48.8\%, 96.0\%] & [79.4\%, 123.2\%] & [97.0\%, 148.7\%] & [76.4\%, 118.8\%] \\
Sports Voucher & [62.1\%, 131.7\%] & [162.8\%, 234.9\%] & [145.9\%, 215.4\%] & [75.2\%, 148.0\%] & [37.6\%, 107.8\%] & [141.4\%, 214.5\%] & [151.5\%, 233.5\%] & [125.9\%, 189.6\%] \\
Market Voucher & [118.1\%, 145.6\%] & [117.9\%, 143.5\%] & [118.8\%, 145.2\%] & [117.1\%, 143.1\%] & [74.2\%, 103.7\%] & [111.1\%, 136.9\%] & [141.6\%, 168.9\%] & [118.6\%, 143.3\%] \\
Agricultural Voucher & 
[84.4\%, 131.5\%] & [71.7\%, 112.9\%] & [82.0\%, 125.0\%] & [67.7\%, 111.4\%] & [18.5\%, 69.0\%] & [60.1\%, 102.7\%] & [98.6\%, 142.3\%] & [76.8\%, 117.0\%] \\
\bottomrule
\end{tabular}
\parbox{26cm}{
\vspace{0.2cm}
Note: The induced consumption rate for the accommodation voucher was
measured based on respondents' answers to the following question:
\emph{``When using the accommodation voucher, did you make any
additional payments beyond the face value of the voucher?"} For the
other five types of vouchers, the induced consumption rate was measured
based on respondents' answers to the following question: \emph{``When
using the \underline{specific voucher}, did you incur any additional
spending beyond the value of the voucher?''} For detailed survey
content, please refer to the appendix. The method for calculating the
induced consumption rate is presented in  \eqref{eq:ic_j} and \eqref{eq:ic}.}
\end{center}
\end{footnotesize}
\end{table}
\end{landscape}

\begin{table}[htbp]
\centering
\begin{footnotesize}
\caption{Estimated Bounds of the Treatment Effect Intensity by Voucher Type}
\label{Tab:tab4}
\begin{tabular}{lcccccc}
\toprule
& Accommodation & Dining & Cultural & Sports & Market & Agricultural \\
\midrule
Lower Bound (5th Percentile)  & 527.85 & 155.87 & 97.50 & 28.55 & 167.22 & 100.35 \\
Upper Bound (95th Percentile) & 860.70 & 164.69 & 117.87 & 116.78 & 182.46 & 140.44 \\
\bottomrule
\end{tabular}
\parbox{15cm}{
\vspace{0.2cm}
Note: All values are expressed in NT\$ millions and are calculated based on Equation~\eqref{eq:it} using the stratified bootstrap procedure.}
\end{footnotesize}
\end{table}

\subsection{Comparison to Previous Studies}

Similar to \citet{xing2023quick}, this paper examines consumer behavioral responses and the policy effectiveness of digital consumption vouchers. \citet{xing2023quick} reported that China's digital voucher policy generated an average of 3.07 yuan in consumer spending for every 1 yuan of government subsidy. This induced consumption effect is substantially stronger than that observed under Taipei Bear Vouchers 2.0 program. A plausible explanation is that the Chinese policy imposed a short, seven-day usage period, which may have heightened consumers' sense of urgency and encouraged more immediate spending behavior.

Further insights can be drawn by comparing the results to Taiwan's 2009 paper-based voucher program. According to \citet{kan2017understanding}, the average MPC under the 2009 policy was approximately 0.25, indicating that each NT\$1 distributed in vouchers led to about NT\$0.24 in unplanned spending. The study also found that consumers who used the vouchers primarily for essential goods exhibited significantly lower MPC and contributed relatively little additional out-of-pocket spending. In contrast, individuals who were younger, more highly educated, more optimistic about the economic outlook, had higher household income, received larger voucher amounts, or experienced recent income growth tended to exhibit higher MPC. Overall, most of the 2009 paper-based vouchers were used for planned purchases. For example, nearly 90\% of spending on essential goods was classified as substitutional, with only 10.6\% attributable to the induced consumption effect. Spending on durable goods and services had an estimated substitution rate of around 75\%.

In contrast, the dining voucher under the Taipei Bear Vouchers 2.0 program, also related to essential consumption, exhibited a significantly lower substitution rate, ranging between 23\% and 38\%. This suggests stronger performance in reducing substitution effects relative to the 2009 program. Moreover, the present study finds that the induced consumption rates for all types of Taipei Bear Vouchers 2.0 exceeded 70\%, indicating a relatively high level of effectiveness in stimulating additional consumer spending.

Differences in consumer responses across the 2009 paper-based voucher program and the Taipei Bear Vouchers 2.0 program may reflect both variations in policy design and underlying consumer characteristics. The 2009 program was implemented nationwide, with participants geographically dispersed and permitted to spend vouchers across a broad range of categories without restriction. In contrast, the Taipei Bear Vouchers 2.0 program was limited to designated spending categories within Taipei City and targeted a more localized population. These structural differences likely shaped consumer spending behavior and, in turn, influenced the effectiveness of the two programs. Another relevant consideration is that our analysis focuses on the treatment effect among individuals who registered for the digital vouchers (i.e., the treated group), and therefore does not account for those who chose not to register (the latent group). If we were to estimate the unconditional treatment effects, i.e., the effects of distributing vouchers to all eligible individuals, the behavioral responses of the latent group could potentially alter the findings. However, given the specific policy context and the objective of assessing the realized impact of the program, our study deliberately concentrates on the treated group, which seeks to measure the effects experienced by voucher registrants and the consequent industry benefits.

\subsection{Economic Impacts of Consumption Vouchers}

To assess the economic impact of Taipei's consumption voucher program, we apply the Taipei Input–Output model to evaluate the effects of the Taipei Bear Vouchers 2.0 policy and to calculate the associated output multipliers based on Equation~\eqref{eq:io}. In this framework, the estimated total induced consumption is treated as a change in final demand, denoted by $\Delta\mathbf{F}$. Specifically, for each voucher type $k$, we compute the induced consumption as the product of the original issued amount and the behavioral adjustment factor $(1 - ES_k)\times(1 + IC_k)$, where $ES_k$ denotes the expenditure substitution rate and $IC_k$ represents the induced consumption rate. These adjusted values are then used as inputs in the input–output model to estimate the broader economic impacts. The coefficient matrix and value-added vector used in Equation~\eqref{eq:io} are provided in Table~\ref{Tab:tabA} in the Appendix.

We first discuss the initial consumption multipliers implied by the overall induced consumption. As shown in Table~\ref{Tab:tab5}, the total initial funding for the six types of vouchers amounted to approximately NT\$584.53 million. Among these, Dining Vouchers accounted for the largest share (70.63\%), followed by Market Vouchers (16.39\%), while Agricultural, Cultural, Sports, and Accommodation Vouchers each comprised less than 5\% of the total budget. By incorporating survey-based estimates of users' expenditure substitution rates and induced consumption rates, we obtain the total induced consumption for each voucher type. For example, the initial budget allocation for Dining Vouchers was NT\$412.85 million. After applying the corresponding substitution and induced consumption effects, the effective consumption is estimated at NT\$453.21 million under the pessimistic scenario and NT\$735.77 million under the optimistic scenario, resulting in multipliers of 1.10 and 1.78, respectively. Overall, after accounting for both substitution and induced consumption effects, the adjusted total induced demand amounts to NT\$689.83 million (pessimistic) and NT\$1,064.96 million (optimistic), which corresponds to 1.18 and 1.82 times the initial funding. Notably, Accommodation Vouchers exhibit the greatest amplification effect, with multipliers of 1.58 and 2.94, due to their relatively low substitution rate and high induced consumption rate.

Given that different types of consumption vouchers are applicable to different categories of consumption, they induce changes in the final demand across various industrial sectors. In our input-output model, the Accommodation Voucher is mapped to the Accommodation Industry; the Dining, Market, and Agricultural Vouchers are mapped to the Retail and Food Services Industry; while the Cultural and Sports Vouchers are mapped to the Arts, Entertainment, and Recreation Services Industry.

\begin{table}
\begin{scriptsize}
\begin{center}
\caption{Estimated Policy Inputs for the Taipei Bear Vouchers 2.0
Program}
\label{Tab:tab5}
\begin{tabular}{lcccccc}
\toprule
& \makecell{Original Amount \\ (NT\$ million)} & \makecell{Percentage \\ of \\ Total} &   \makecell{Induced Demand\\ (NT\$ million) \\ pessimistic} &\makecell{Induced Demand\\ (NT\$ million) \\ optimistic} &  \makecell{Multiplier\\pessimistic} & \makecell{Multiplier\\optimistic} \\
\cmidrule(l){2-7}
Accommodation Voucher & 11.28 & 1.93\% & 17.84 & 33.17  &1.58 & 2.94\\
Dining Voucher        & 412.85 & 70.63\% & 453.21 & 735.77  & 1.10  & 1.78\\
Cultural Voucher      & 29.04 & 4.97\% & 33.53 & 49.30 & 1.15 & 1.70\\
Sports Voucher        & 14.52 & 2.48\% & 11.45 & 23.69  & 0.79 & 1.63\\
Market Voucher        & 95.80 & 16.39\% & 144.92 & 187.05  & 1.51 & 1.95\\
Agricultural Voucher  & 21.04 & 3.60\% & 28.88 & 35.98  & 1.37 & 1.71\\
Total                 & 584.53 & 100.00\% &689.83 & 1,064.96  & 1.18 & 1.82\\
\bottomrule
\end{tabular}
\parbox{17cm}{
\vspace{0.2cm}
Note: The original issued amounts are provided by the Taipei City Government, and these figures include the second-round extra stimulus vouchers. The optimistic and pessimistic estimates of the induced change in demand are calculated using the expression $(1 - ES_k) \times (1 + IC_k)$, where $ES_k$ denotes the substitution effect and $IC_k$ denotes the induced consumption effect for voucher type $k$. These estimates are obtained by multiplying the issued amount by the above expression, using the mean values from the lower and upper bounds of the respective confidence intervals for substitution and induced consumption effects. The monetary values in the table are expressed in 2022 price levels.
}
\end{center}
\end{scriptsize}
\end{table}

Table~\ref{Tab:tab6} presents the estimated economic benefits of the Taipei Bear Vouchers 2.0 program on Taipei City's industrial economy under three scenarios. The baseline scenario reflects the economic impact based solely on the original budget allocation, treating it as a direct change in final demand without accounting for expenditure substitution or induced consumption effects. This scenario assumes a total policy input of NT\$584.53 million. In addition to the baseline, the table includes a pessimistic scenario and an optimistic scenario, both of which incorporate behaviorally adjusted estimates of final demand based on substitution and induced consumption effects as shown in Table~\ref{Tab:tab5}.\footnote{Since the Taipei City input–output table used in this paper is based on the year 2016, the estimated final demand values for each type of Taipei Bear Vouchers 2.0 under different scenarios in 2022 were first deflated to 2016 price levels using the Consumer Price Index (CPI) before conducting the input–output analysis. Subsequently, the estimated economic benefits derived from the input–output analysis were re-inflated to 2022 monetary values using the CPI as the inflation adjustment index.}

As shown in Table~\ref{Tab:tab6}, when consumer behavioral responses are excluded, the voucher policy is estimated to increase Taipei's GDP by approximately NT\$566.32 million. After accounting for substitution and induced consumption effects, the estimated GDP impact rises to NT\$668.18 million in the pessimistic case and NT\$1,029.83 million in the optimistic case. These adjustments correspond to additional gains of NT\$101.85 million and NT\$463.51 million, respectively. Furthermore, the output multiplier improves from 0.969 in the baseline to 1.762 in the optimistic case, indicating a significant enhancement in the policy's effectiveness when actual consumer behavior is taken into consideration.

At the industry level, the Retail Trade and Food Services sector exhibits the largest increase in output. Its contribution to GDP rises from NT\$397.13 million in the baseline scenario to NT\$470.08 million under the pessimistic adjustment and NT\$718.96 million under the optimistic case. The additional gains in this sector range from NT\$72.96 million to NT\$321.84 million, reflecting its dominant role in absorbing consumer spending and its large share of the overall policy allocation.

The Arts, Entertainment, and Recreation Services sector also experiences meaningful growth, with GDP rising from NT\$30.45 million in the baseline to NT\$31.62 million and NT\$51.19 million in the pessimistic and optimistic cases, respectively. This sector's output expands by up to NT\$20.74 million depending on consumer response intensity. Similarly, the Accommodation sector sees GDP increase from NT\$5.58 million to NT\$8.76 million and NT\$16.24 million, with corresponding gains of NT\$3.19 million and NT\$10.66 million.

Sectors not directly targeted by the vouchers also show notable output increases due to inter-industry linkages. For example, the Finance, Legal, Real Estate, and Professional Services sector records GDP growth from NT\$72.98 million in the baseline to NT\$86.45 million and NT\$133.28 million in the adjusted scenarios. These figures represent additional output of NT\$13.47 million in the pessimistic case and NT\$60.30 million in the optimistic case.

In summary, the input–output analysis emphasizes the critical role of consumer behavior in evaluating the effectiveness of consumption-based stimulus programs. When substitution and induced consumption effects are not taken into account, the estimated policy impact may be considerably understated. Incorporating these effects provides a more accurate and comprehensive view of both the direct and indirect benefits across industries.

\begin{table}
\begin{footnotesize}
\begin{center}
\caption{Economic Benefit of Taipei Bear Vouchers 2.0 Program}
\label{Tab:tab6}
\begin{tabular}{lccccc}
\toprule
Industry Sector & Baseline & Pessimistic & Optimistic & \makecell{Difference\\(Pessimistic)} & \makecell{Difference\\(Optimistic)}\\
\midrule
\makecell[l]{Agriculture, Forestry, Fishery, and \\ Animal Husbandry} & 0.052 & 0.062 & 0.095 & 0.010 & 0.043 \\
Mining & 0.010 & 0.011 & 0.017 & 0.002 & 0.008 \\
\makecell[l]{Light Manufacturing (Food, Textiles, Wood, \\ Paper, Printing)} & 6.866 & 8.129 & 12.488 & 1.263 & 5.621 \\
\makecell[l]{Chemical, Petrochemical, and Rubber and \\ Plastics Manufacturing} &1.251 & 1.493 & 2.311 & 0.243 & 1.060 \\
\makecell[l]{Metal and Non-metallic Mineral \\ Products Manufacturing} & 0.139 & 0.166 & 0.257 & 0.027 & 0.118 \\
\makecell[l]{Electronics, Electrical Machinery, and \\ Computer Optical Products
Manufacturing} & 0.181 & 0.210 & 0.326 & 0.029 & 0.146 \\
\makecell[l]{Machinery, Transportation Equipment, \\ Furniture, and Electrical
Equipment} & 0.184 & 0.214 & 0.334 & 0.030 & 0.150 \\
Utilities and Waste Management & 7.995 & 9.508 & 14.710 & 1.513 & 6.715 \\
Construction & 2.498 & 2.960 & 4.573 & 0.462 & 2.075 \\
Wholesale Trade & 15.515 & 18.365 & 28.237 & 2.850 & 12.723 \\
Retail Trade and Food Services & 397.126 & 470.081 & 718.964 & 72.955 & 321.838 \\
Transportation, Storage, and Logistics & 3.878 & 4.595 & 7.075 & 0.717 & 3.197 \\
Accommodation & 5.575 & 8.762 & 16.238 & 3.187 & 10.663 \\
\makecell[l]{Music, Publishing, and Information \\ Technology Services} & 9.661 & 11.386 & 17.751 & 1.725 & 8.090 \\
\makecell[l]{Finance, Legal, Real Estate, and \\ Professional Services (Design,
etc.)} & 72.981 & 86.448 & 133.284 & 13.466 & 60.303 \\
\makecell[l]{Employment Agencies, Travel Agencies, \\ Security, Administration, and \\
Defense Services} & 9.244 & 10.911 & 16.889 & 1.667 & 7.645 \\
\makecell[l]{Education, Medical, and \\ Social Work Services} & 0.189 & 0.223 & 0.346 & 0.034 & 0.157 \\
\makecell[l]{Arts, Entertainment, and \\ Recreation Services} & 30.449 & 31.619 & 51.189 & 1.170 & 20.740 \\
Other Services & 2.529 & 3.031 & 4.749 & 0.502 & 2.220 \\
\midrule
Total & 566.323 & 668.175 & 1029.833 &  & \\
Output Multiplier  & 0.969 & 1.143 & 1.762 & & \\
\bottomrule
\end{tabular}
\parbox{16cm}{
\vspace{0.2cm}
Note: The output multiplier is defined as the ratio of the change in
gross domestic product (GDP) to the original amount of policy
expenditure. The original amount of policy expenditure is NT\$584.53
million. All units are in measured in million except the output multiplier. The monetary values in the table are expressed in 2022 price levels.
}
\end{center}
\end{footnotesize}
\end{table}

\section{Conclusion and Policy Implications}
During the COVID-19 pandemic, governments worldwide adopted expansionary fiscal policies to stabilize domestic demand and support economic recovery. Among the tools implemented, direct cash transfers and consumption vouchers emerged as two common approaches. While cash transfers are administratively simple and offer recipients greater flexibility, empirical research has found that their marginal propensity to consume is often limited. Many households use these funds for savings or debt repayment, reducing their short-term stimulative impact. Moreover, cash transfers are generally not directed at specific sectors, making it difficult to target support to the industries most affected by economic downturns.

By contrast, consumption vouchers are more restrictive in scope but can be designed to promote spending in targeted categories of goods and services. Features such as designated usage and expiration dates help accelerate consumption and guide spending toward specific sectors, increasing the overall output multiplier.

This study evaluates the economic impact of the Taipei Bear Vouchers 2.0 program, a digital voucher scheme implemented by the Taipei City Government in 2022. Using verified first-hand user data from the TaipeiPASS system and a regional input–output model, we examine how actual consumer behavior influences policy effectiveness. Our analysis focuses on three key behavioral mechanisms: expenditure substitution, induced consumption, and treatment intensity across voucher types. These parameters allow for an adjustment of initial budget allocations to more accurately reflect the policy's true economic contribution.

The results reveal clear differences across voucher types. Accommodation vouchers generate the lowest substitution and strongest induced spending, making them particularly effective in stimulating incremental consumption. Even modest increases in voucher face value lead to additional marginal spending, especially when such increases are unanticipated or supported by merchant-side promotions. Input–output simulations show that when behavioral responses are incorporated, the estimated economic benefits of the voucher program increase significantly relative to estimates based only on nominal allocations. In addition to direct stimulus, the program also induces output gains in untargeted sectors through inter-industry linkages.

Based on these findings, we draw four key policy implications. First, consumer behavior is central to determining the effectiveness of voucher-based stimulus programs. Second, compared to direct cash transfers, well-designed vouchers can yield stronger consumption inducement and higher multipliers, particularly when targeting specific sectors. Third, program design should emphasize allocating vouchers to sectors with lower substitution and higher induced consumption effects to maximize policy efficiency. Fourth, expanding voucher eligibility to include non-local consumers can help attract additional spending and amplify regional economic activity.

\bibliographystyle{ecta}
\bibliography{voucher}

\newpage
\textbf{Appendix: Survey Questions on the Expenditure Substitution
Effect and the Induced Consumption Effect}

\noindent
\textbf{A.1 Demographic Variables}

\begin{enumerate}
\item
  \textbf{Age:}
  \begin{enumerate}
\item
  \emph{Under 20 years old}
\item
  \emph{20--29 years old}
\item
  \emph{30--39 years old}
\item
  \emph{40--49 years old}
\item
  \emph{50--59 years old}
\item
  \emph{60 years old or above}
\end{enumerate}

\item
  \textbf{Gender:}
  \begin{enumerate}
  \item 
  \emph{Male}
\item 
\emph{Female}
  \end{enumerate}

\item
  \textbf{Residence:}
\begin{enumerate}
\item
  \emph{Taipei City}
\item
  \emph{New Taipei City, Keelung City, or Taoyuan City}
\item
  \emph{Other cities/counties in Taiwan}
\end{enumerate}

\end{enumerate}

\noindent
\textbf{A.2 Expenditure Substitution Effect}

\begin{enumerate}
\item
  \textbf{The Accommodation vouchers}: ``\emph{Did you make this
  accommodation consumption because you received the accommodation
  voucher?'' (a) Yes; (b) No}
\item
  \textbf{The Dining vouchers}: ``\emph{Did you make this consumption
  because you received the dining voucher?'' (a) Yes; (b) No}
\item
  \textbf{The Cultural vouchers:} ``\emph{Did you make this consumption
  because you received the cultural voucher?'' (a) Yes; (b) No}
\item
  \textbf{The Sports vouchers:} ``Did you make this consumption because
  you received the sports voucher?'' \emph{(a) Yes; (b) No}
\item
  \textbf{The Market vouchers:} ``Did you make this consumption because
  you received the market voucher?'' \emph{(a) Yes; (b) No}
\item
  \textbf{The Agricultural vouchers:} ``Did you make this consumption
  because you received the agricultural voucher?'' \emph{(a) Yes; (b)
  No}
\end{enumerate}

\noindent
\textbf{A3 Survey Questions on Induced Consumption Effect}

\begin{enumerate}
\item
  \textbf{The Accommodation vouchers}: \emph{``When using the
  accommodation voucher, did you make any additional payments beyond the
  face value of the voucher?"}
  \begin{enumerate}
\item
  No additional spending
\item
  \emph{NT\$1--1,000}
\item
  \emph{NT\$1,001--3,000}
\item
  \emph{NT\$3,001--5,000}
\item
  \emph{NT\$5,001--8,000}
\item
  \emph{NT\$8,001--10,000}
\item
  \emph{NT\$10,001--20,000}
\item
  \emph{More than NT\$20,001}
\end{enumerate}

\item
  \textbf{The Dining vouchers}: \emph{``When using the dining voucher,
  did you incur any additional spending beyond the value of the
  voucher?"}
\begin{enumerate}
\item
  \emph{No additional spending}
\item
  \emph{NT\$1--50}
\item
  \emph{NT\$51--100}
\item
  \emph{NT\$101--250}
\item
  \emph{NT\$251--500}
\item
  \emph{NT\$501--1,000}
\item
  \emph{NT\$1,001--2,000}
\item
  \emph{More than NT\$2,001}
\end{enumerate}

\item
  \textbf{The Cultural vouchers:} \emph{"When using the cultural voucher
  did you incur any additional spending beyond the value of the
  voucher?"}
  \begin{enumerate}
  \item
    \emph{No additional spending}
  \item
    \emph{NT\$1--50}
  \item
    \emph{NT\$51--100}
  \item
    \emph{NT\$101--250}
  \item
    \emph{NT\$251--500}
  \item
    \emph{NT\$501--1,000}
  \item
    \emph{NT\$1,001--2,000}
  \item
    \emph{More than NT\$2,001}
  \end{enumerate}

\item
  \textbf{The Sports vouchers:} \emph{"When using the sports voucher,
  did you incur any additional spending beyond the value of the
  voucher?"}

\begin{enumerate}
\item
  \emph{No additional spending}
\item
  \emph{NT\$1--50}
\item
  \emph{NT\$51--100}
\item
  \emph{NT\$101--250}
\item
  \emph{NT\$251--500}
\item
  \emph{NT\$501--1,000}
\item
  \emph{NT\$1,001--2,000}
\item
  \emph{More than NT\$2,001}
\end{enumerate}

\item
  \textbf{The Market vouchers:} \emph{"When using the market voucher,
  did you incur any additional spending beyond the value of the
  voucher?"}
  \begin{enumerate}
\item
  \emph{No additional spending}
\item
  \emph{NT\$1--50}
\item
  \emph{NT\$51--100}
\item
  \emph{NT\$101--250}
\item
  \emph{NT\$251--500}
\item
  \emph{NT\$501--1,000}
\item
  \emph{NT\$1,001--2,000}
\item
  \emph{More than NT\$2,001}
\end{enumerate}

\item
  \textbf{The Agricultural vouchers:} \emph{"When using the agricultural
  voucher,} \emph{did you incur any additional spending beyond the value
  of the voucher?}

  \begin{enumerate}
\item
  \emph{No additional spending}
\item
  \emph{NT\$1--50}
\item
  \emph{NT\$51--100}
\item
  \emph{NT\$101--250}
\item
  \emph{NT\$251--500}
\item
  \emph{NT\$501--1,000}
\item
  \emph{NT\$1,001--2,000}
\item
  \emph{More than NT\$2,001}
\end{enumerate}

\end{enumerate}

\noindent
\textbf{A4 Input-Output Coefficients Matrix and Added Value}

\begin{landscape}
\begin{table}
\begin{tiny}
\begin{center}
\caption{Input-Output Coefficients Matrix and Added Value}
\label{Tab:tabA}
\begin{tabular}{|l|c|c|c|c|c|c|c|c|c|c|c|c|c|c|c|c|c|c|c|}
\hline                                                                                             
\makecell[l]{Agriculture, Forestry, Fishery, and \\ Animal Husbandry} & 1.001 & 0.000 & 0.001 & 0.000 & 0.000 & 0.000 & 0.000 & 0.000 & 0.000 & 0.000 & 0.000 & 0.000 & 0.000 & 0.000 & 0.000 & 0.000 & 0.000 & 0.000 & 0.000 \\
\hline
Mining & 0.000 & 1.000 & 0.000 & 0.000 & 0.001 & 0.000 & 0.000 & 0.000 & 0.002 & 0.000 & 0.000 & 0.000 & 0.000 & 0.000 & 0.000 & 0.000 & 0.000 & 0.000 & 0.000 \\
\hline
\makecell[l]{Light Manufacturing (Food, Textiles, Wood, \\ Paper, Printing)} & 0.053 & 0.002 & 1.081 & 0.003 & 0.003 & 0.001 & 0.006 & 0.001 & 0.011 & 0.006 & 0.048 & 0.002 & 0.025 & 0.014 & 0.007 & 0.007 & 0.006 & 0.017 & 0.008 \\
\hline
\makecell[l]{Chemical, Petrochemical, and Rubber and \\ Plastics Manufacturing} & 0.014 & 0.028 & 0.023 & 1.096 & 0.013 & 0.007 & 0.013 & 0.014 & 0.013 & 0.007 & 0.010 & 0.031 & 0.018 & 0.003 & 0.004 & 0.006 & 0.006 & 0.004 & 0.014 \\
\hline
\makecell[l]{Metal and Non-metallic Mineral \\ Products Manufacturing} & 0.001 & 0.001 & 0.002 & 0.001 & 1.037 & 0.002 & 0.014 & 0.001 & 0.029 & 0.000 & 0.001 & 0.001 & 0.002 & 0.001 & 0.001 & 0.001 & 0.001 & 0.001 & 0.001 \\
\hline
\makecell[l]{Electronics, Electrical Machinery, and \\ Computer Optical Products Manufacturing} & 0.000 & 0.001 & 0.000 & 0.000 & 0.000 & 1.054 & 0.004 & 0.001 & 0.004 & 0.001 & 0.001 & 0.000 & 0.001 & 0.007 & 0.004 & 0.001 & 0.001 & 0.003 & 0.001 \\
\hline
\makecell[l]{Machinery, Transportation Equipment, \\ Furniture, and Electrical Equipment} & 0.001 & 0.007 & 0.001 & 0.001 & 0.002 & 0.001 & 1.022 & 0.004 & 0.011 & 0.001 & 0.001 & 0.005 & 0.002 & 0.002 & 0.001 & 0.001 & 0.002 & 0.003 & 0.010 \\
\hline
Utilities and Waste Management & 0.010 & 0.009 & 0.031 & 0.034 & 0.039 & 0.014 & 0.011 & 1.114 & 0.008 & 0.011 & 0.034 & 0.011 & 0.055 & 0.013 & 0.007 & 0.023 & 0.022 & 0.021 & 0.015 \\
\hline
Construction & 0.004 & 0.014 & 0.005 & 0.004 & 0.007 & 0.002 & 0.004 & 0.016 & 1.007 & 0.010 & 0.014 & 0.009 & 0.018 & 0.017 & 0.032 & 0.010 & 0.009 & 0.011 & 0.005 \\
\hline
Wholesale Trade & 0.044 & 0.027 & 0.067 & 0.030 & 0.059 & 0.024 & 0.069 & 0.019 & 0.073 & 1.007 & 0.039 & 0.019 & 0.025 & 0.024 & 0.013 & 0.011 & 0.038 & 0.017 & 0.024 \\
\hline
Retail Trade and Food Services & 0.034 & 0.023 & 0.029 & 0.011 & 0.022 & 0.005 & 0.033 & 0.007 & 0.036 & 0.017 & 1.022 & 0.018 & 0.020 & 0.017 & 0.011 & 0.019 & 0.057 & 0.016 & 0.035 \\
\hline
Transportation, Storage, and Logistics & 0.011 & 0.057 & 0.016 & 0.009 & 0.019 & 0.005 & 0.014 & 0.007 & 0.031 & 0.024 & 0.016 & 1.073 & 0.013 & 0.012 & 0.009 & 0.011 & 0.011 & 0.008 & 0.010 \\
\hline
Accommodation & 0.000 & 0.000 & 0.000 & 0.000 & 0.000 & 0.000 & 0.001 & 0.001 & 0.000 & 0.002 & 0.001 & 0.000 & 1.000 & 0.000 & 0.000 & 0.000 & 0.000 & 0.000 & 0.000 \\
\hline
\makecell[l]{Music, Publishing, and Information \\ Technology Services} & 0.003 & 0.006 & 0.008 & 0.004 & 0.005 & 0.002 & 0.006 & 0.003 & 0.010 & 0.021 & 0.027 & 0.015 & 0.064 & 1.164 & 0.030 & 0.028 & 0.018 & 0.057 & 0.024 \\
\hline
\makecell[l]{Finance, Legal, Real Estate, and \\ Professional Services (Design, etc.)} & 0.024 & 0.065 & 0.051 & 0.025 & 0.034 & 0.020 & 0.037 & 0.023 & 0.063 & 0.098 & 0.184 & 0.065 & 0.184 & 0.093 & 1.162 & 0.067 & 0.034 & 0.113 & 0.066 \\
\hline
\makecell[l]{Employment Agencies, Travel Agencies, \\ Security, Administration, and \\ Defense Services} & 0.005 & 0.013 & 0.011 & 0.007 & 0.012 & 0.005 & 0.012 & 0.013 & 0.015 & 0.026 & 0.022 & 0.029 & 0.029 & 0.018 & 0.016 & 1.018 & 0.014 & 0.027 & 0.008 \\
\hline
\makecell[l]{Education, Medical, and \\ Social Work Services} & 0.000 & 0.000 & 0.000 & 0.000 & 0.001 & 0.000 & 0.000 & 0.001 & 0.001 & 0.001 & 0.000 & 0.001 & 0.001 & 0.003 & 0.001 & 0.005 & 1.001 & 0.001 & 0.000 \\
\hline
\makecell[l]{Arts, Entertainment, and \\ Recreation Services}& 0.000 & 0.001 & 0.001 & 0.000 & 0.001 & 0.000 & 0.001 & 0.000 & 0.001 & 0.003 & 0.003 & 0.001 & 0.002 & 0.006 & 0.004 & 0.002 & 0.001 & 1.043 & 0.010 \\
\hline
Other Services & 0.001 & 0.011 & 0.004 & 0.002 & 0.005 & 0.002 & 0.003 & 0.002 & 0.007 & 0.010 & 0.006 & 0.016 & 0.024 & 0.004 & 0.005 & 0.007 & 0.004 & 0.008 & 1.015 \\
\hline
\hline
Added Value & 0.522 & 0.580 & 0.261 & 0.222 & 0.242 & 0.327 & 0.256 & 0.413 & 0.305 & 0.716 & 0.732 & 0.439 & 0.482 & 0.550 & 0.698 & 0.715 & 0.753 & 0.645 & 0.649 \\
\hline
\end{tabular}
\end{center}
\end{tiny}
\end{table}
\end{landscape}

\noindent
\textbf{A5 Method for Compiling the Taipei City Input--Output Table}

This paper adopts the non-survey approach, specifically the Simple Location Quotient (SLQ) method, to construct the input--output table for Taipei City. This method adjusts the national input coefficients to reflect the economic structure of a specific region. Originally proposed by  \citet{Walter1953}, the approach has been widely applied in the construction of regional input--output tables \citep{miller2009input}.

Based on Taiwan's 2016 national input--output table, this paper incorporates industrial output statistics for Taipei City, obtained from Taiwan's \textit{Industrial and Service Census Report}, to calculate sectoral location quotients (SLQs). These SLQs are then used to adjust the national input coefficients and estimate a region-specific technical coefficient matrix for Taipei City. The detailed methodology and technical procedures are described below.

Let $X_i^r$ denote the output of industry $i$ in region $r$, and $X_i$ represent the national output of industry $i$. The SLQ for industry $i$ in region $r$ is defined as:
\begin{equation}
\text{SLQ}_i^r = \left( \frac{X_i^r}{\sum_{i=1}^n X_i^r} \right) \Big/ \left( \frac{X_i}{\sum_{i=1}^n X_i} \right).
\tag{A.1}
\end{equation}
The value of $\text{SLQ}_i^r$ indicates the ratio of industry $i$'s share of total output in region $r$ to its corresponding share in the national economy. If $\text{SLQ}_i^r > 1$, industry $i$ is more concentrated in region $r$ than the national average. Conversely, if $\text{SLQ}_i^r < 1$, the industry's share in region $r$ is smaller than the national average.

When $\text{SLQ}_i^r > 1$, it is assumed that the output of industry $i$ in region $r$ exceeds local demand; therefore, the regional input coefficients for industry $i$ are set equal to the national coefficients. In contrast, if $\text{SLQ}_i^r < 1$, it is inferred that the regional output of industry $i$ cannot fully meet local demand. In this case, the national input coefficients are adjusted by the SLQ to estimate the regional input coefficients, as shown below:
\begin{equation}
a_{ij}^{rr} =
\begin{cases}
a_{ij} \cdot \text{SLQ}_i^r, & \text{if } \text{SLQ}_i^r < 1 \\
a_{ij}, & \text{if } \text{SLQ}_i^r \geq 1,
\end{cases}
\tag{A.2}
\end{equation}
where $a_{ij}^{rr}$ represents the regional input coefficient for industry $i$ using input from industry $j$, and $a_{ij}$ denotes the corresponding national input coefficient. The matrix composed of $a_{ij}^{rr}$ values constitutes the coefficient matrix $\MBA$ used in this paper.

\end{document}